\documentstyle[12pt,epsf]{article}
\newcommand{\eqnum}[1]{\setcounter{equation}{#1}}

\newcommand{\square}{\kern1pt\vbox{\hrule height
1.2pt\hbox{\vrule width 1.2pt\hskip 3pt
   \vbox{\vskip 6pt}\hskip 3pt\vrule width 0.6pt}\hrule
height 0.6pt}\kern1pt}
\newcommand{\lsim}{\mbox{$\;\lower-.2ex
\hbox{$\textstyle<$}\;\!\!\!\!\!\!
\lower.7ex\hbox{$\textstyle \sim$}\;$}}
\newcommand{\gsim}{\mbox{$\;\lower-.2ex
\hbox{$\textstyle>$}\;\!\!\!\!\!\!
\lower.7ex\hbox{$\textstyle \sim$}\;$}}
\newcommand{\rh}{r_{\scriptscriptstyle H}}
\newcommand{\OmH}{\Omega_{H}}
\newcommand{\PhiH}{\Phi_{H}}
\newcommand{\hspc}{\hspace*{\parindent}}
\newcommand{\Qmax}{Q_{\mbox{\scriptsize max}}}
\newcommand{\Qex}{Q_{\mbox{\scriptsize ex}}}
\newcommand{\ri}{\rho_{1}^*}
\newcommand{\rf}{\rho_{2}^*}
\newenvironment{figcaption}[2]{
 \vspace{0.3cm}
 \refstepcounter{figure}
 \label{#1}
 \begin{center}
 \begin{minipage}{#2}
 \begingroup \small Fig. \thefigure: }{
 \endgroup
 \end{minipage}
 \end{center}}
\newcommand{\singlefig}[2]{
\begin{center}
\begin{minipage}{#1}
\epsfxsize=#1
\epsffile{#2}
\end{minipage}
\end{center}}
\newcommand{\segmentfig}[3]{
\begin{minipage}{#1}
\epsfxsize=#1
\epsffile{#2}
\begin{center}
{\small \mbox{#3}}
\end{center}
\end{minipage}}

\begin{document}

\begin{titlepage}
\baselineskip .15in
\begin{flushright}
WU-AP/46/95
\end{flushright}

{}~\\

\vskip 1.5cm
\begin{center}
{\bf

\vskip 1.5cm
{\large Evaporation and Fate of Dilatonic Black Holes}

}\vskip .8in

{\sc  Jun-ichirou Koga}$^{(a)}$  and
{\sc  Kei-ichi Maeda}$^{(b)}$\\[1em]
 {\em Department of Physics, Waseda University,
Shinjuku-ku, Tokyo 169, Japan}

\end{center}
\vfill
 \begin{abstract}
We study both spherically symmetric and rotating
 black holes with dilaton coupling,
 and discuss  the evaporation of these black holes
via Hawking's quantum radiation and their fates.
We find that the dilaton coupling constant
$\alpha$ drastically affects
the emission rates, and therefore the fates of the black holes.
When the charge is conserved, the emission rate from
the non-rotating hole is drastically changed beyond $\alpha =
1$ (a superstring theory) and diverges in the
extreme limit. In the rotating cases, we analyze the
slowly
 rotating black hole solution with arbitrary
$\alpha$ as well as  three exact solutions, the Kerr--Newman
($\alpha = 0$), and Kaluza--Klein ($\alpha = \sqrt{3}$), and
Sen black hole ($\alpha = 1$ and with axion field). Beyond
the same critical value of $\alpha
\sim 1$, the emission rate becomes very large near the
maximally charged limit, while for $\alpha<1$ it remains finite. The
black hole with $\alpha > 1$  may evolve into a naked
singularity due to its large emission rate. We also
consider the effects of a discharge process  by
investigating superradiance for the non-rotating dilatonic
black hole.
 \end{abstract}

\vfill
\begin{center}
August, 1995
\end{center}
\vfill
(a)~~electronic mail : 695L5084@cfi.waseda.ac.jp\\
(b)~~electronic mail :  maeda@cfi.waseda.ac.jp\\
\end{titlepage}

\baselineskip .3in

\vspace{.5 cm}
\normalsize
\baselineskip = 24pt
\renewcommand{\large}{\normalsize}
\renewcommand{\Large}{\normalsize}
\renewcommand{\huge}{\normalsize}
\section{Introduction}
\eqnum{0}
\hspc
The unification of all fundamental interactions including
gravity is one of the final goals of theoretical physics.
The electromagnetic
 and weak interactions were unified by Weinberg and
Salam, and grand unified theories (GUTs) have been proposed
as a unification model of three fundamental interactions. In
this unification scheme, all interactions are described
by gauge fields. Furthermore, supersymmetry is
proposed to unify interaction (bosons) and matter
(fermions), and gravity could be included in  a
supergravity theory. Such unified theories are  sometimes
discussed in higher-dimensions.  Then, the idea of
superstring  arises as an approach to the unification of  all
interactions and particles, giving ``theory of everything".

Then we have recognized  that gravity is one of the most
important keys for unification.  In order to understand
the role of gravity in a fundamental unified theory, it
is necessary and helpful to study concrete
physical phenomena with  strong gravity such as cosmology or
black holes.  We find new aspects of gravity and other
fundamental fields through such theoretical studies, which might give
us hints about unification.

In the
effective theories derived from the higher-dimensional unified
theories\cite{Callan}, the dilaton field couples
to other known matter fields. The
coupling constant  depends on the fundamental unified
theory and the dimensionality of spacetime. Thus it is
important to study how the coupling affects physical
phenomena.
The coupling plays some
important roles  in
 black hole physics\cite{GM} as well as in
cosmology\cite{Dcos}.
In this paper,  we  further study
 effects of a dilaton
field on black hole physics, and in particular we will analyze
the role of Hawking's quantum radiation.

We  consider the
model  with a dilaton field coupled to a U(1) gauge field,
 i.e., the Einstein--Maxwell--dilaton theory. The action is
\begin{equation}
S = \frac{1}{16 \pi} \int \mbox{d}^{4} x \sqrt{- g}
\left[ R - 2 \left( \nabla \phi \right)^{2} - \mbox{e}^{-
2 \alpha \phi} F^{2} \right] \;  ,
 \label{eqn:dilaction}
\end{equation}
where
$\phi$ and $F_{\mu\nu}$ are  a dilaton field and  U(1)
gauge field, respectively, with coupling
constant $\alpha$\cite{unit}. For a superstring, we may also
include an  axion field $H_{\mu\nu\rho}$.  The action
is then
\begin{equation}
 S = \frac{1}{16 \pi} \int \mbox{d}^{4} x \sqrt{- g}
\left[ R - 2 \left( \nabla \phi \right)^{2} - \mbox{e}^{-
2 \phi} F^{2} - \frac{1}{12} \mbox{e}^{- 4 \phi} H^2
\right] \;  .
 \label{eqn:senaction}
\end{equation}

The
action (\ref{eqn:dilaction}) reduces to the
Einstein--Maxwell theory when the coupling constant $\alpha
= 0$. The black hole solution for this case is the  well
known Kerr--Newman family. The case of
$\alpha = \sqrt{3}$
corresponds to the 4-dimensional effective
model reduced from the 5-dimensional Kaluza--Klein
 theory.
The
action (\ref{eqn:senaction}), in which the dilaton coupling
constant $\alpha$ to the U(1) gauge field is unity, is a
bosonic part of the low energy limit of superstring theory.

 The exact spherically-symmetric  dilatonic
black hole solution with arbitrary coupling
constant $\alpha$ is known\cite{GM,GHS}. They have some interesting
thermodynamical properties, which are not found in the
conventional charged (Reissner-Nordstr\"{o}m) black hole. In
particular, the temperature of the black hole in the extreme
limit depends drastically on $\alpha$. If $\alpha < 1$, the
temperature of the black hole vanishes in the extreme limit, as
does that of the Reissner-Nordstr\"om black hole. On the other
hand, the temperature of the extreme  black hole with
$\alpha > 1$ diverges.  For $\alpha = 1$,
it is a non-zero finite value. This new thermodynamical
property implies that the emission rate of Hawking
quantum radiation may be completely different, depending on
the coupling constant. We expect that when $\alpha > 1$,
 the emission rate diverges in the extreme limit, because the
temperature diverges.
The black hole  may evaporate very rapidly.
However, it was pointed
out \cite{HW} that for $\alpha >1$, the effective potential,
 over which created particles travel to
an asymptotically flat region to evaporate, grows infinitely
high in the extreme limit.  Hence, Holzhey and Wilczek
expected that the emission rate will be  suppressed to a
finite value. Since these two features are competing processes in
Hawking radiation,  it is not trivial to decide whether  or not
the emission rate from the extreme dilatonic black holes
with $\alpha > 1$ diverges. Thus, we analyze the emission rates numerically
under the assumption that the charge is conserved, and
clarify what happens in the extreme limit.  This is the main
purpose of the present paper.

 In
addition to the spherically symmetric black hole,
 rotating dilatonic black holes
also have similar thermodynamical
properties\cite{HH,Sen,KM}. We
 considered superradiance around the rotating dilatonic
black holes in the previous paper\cite{KM} and showed that
there is a critical value ($\alpha \sim 1$) beyond which the
emission rate changes drastically. In this paper, we extend
our analysis to include the role of the temperature, i.e.,
Hawking quantum radiation, which automatically includes a
superradiant effect,  and discuss the fate of rotating black
holes due to the evaporation process. We only know two exact rotating
black hole solutions for the
action (\ref{eqn:dilaction}): Kerr--Newman
($\alpha = 0$) and Kaluza--Klein solution ($\alpha =
\sqrt{3}$)\cite{FZB}. In the superstring case ($\alpha =
1$), Sen\cite{Sen} derived a rotating black hole solution
for the action(\ref{eqn:senaction}). This solution is not
exactly  the same as those in the model
(\ref{eqn:dilaction}), but we
expect that the existence of the axion field will not
drastically change the dependence of the emission rate
 on the dilaton
coupling. Hence, we   analyze
these three black hole solutions and compare their emission rates.
 Besides these exact solutions, we
consider an approximate solution of slowly rotating black
holes with arbitrary coupling
$\alpha$\cite{HH,Shiraishi1}.

 All rotating  dilatonic black holes
reduce to the  Kerr solution when their charges vanish.
We expect that  the coupling constant dependence is
most noticeable when the black hole is highly  charged.
We therefore analyze Hawking radiation from highly
charged black holes.
As we know, a charged black hole generally emits its
charge at a high rate  in the process of evaporation and so
its charge will be quickly lost, unless the charge is
conserved. Because we are now interested in
the effect of the dilaton coupling on the emission rates,
we first assume that the  charge is conserved, which is
true for a central charge.
We then study the discharge processes to see
how it is affected by the dilaton coupling.

 This paper is organized as follows. In the next
section, we study Hawking radiation for a
spherically symmetric dilatonic black hole and analyze the
behaviour of the emission rate in  the extreme limit. The
emission rates from rotating black holes  are
presented
in the section 3. It is assumed  that the charge
of the black hole is conserved. We discuss  the
evolution and fate  of these black holes.
The effects of the discharge process are considered by calculating
superradiance in the spherically symmetric black hole
in the section 4. Finally, we give  our conclusions and
remarks in the final section.

\section{Hawking Radiation from Spherically Symmetric
 Dilatonic Black Holes}
\eqnum{0}
\hspc
We first consider spherically symmetric dilatonic black holes.
In this case we know the exact solution with arbitrary
 coupling constant $\alpha$ \cite{GM,GHS}, which is given by
\begin{eqnarray}
\mbox{d}s^{2} & = & -{\Delta(\rho) \over {R^2(\rho)}}
\: \mbox{d}t^{2} + \frac{{R^2(\rho)}}{\Delta(\rho)} \: \mbox{d}\rho^{2} +
{ {R}^{2}(\rho)} (\mbox{d} \theta^{2} + \sin^{2}\theta \:
\mbox{d} \varphi^{2} )  \: , \nonumber \\
A_{t} & = & \frac{Q}{\rho}, \; \; \; \phi = \frac{\alpha}{1+\alpha^{2}}
\ln \left(1 - \frac{\rho_{-}}{\rho} \right) \; ,   \label{eqn:nonrotsol}
\end{eqnarray}
where
\begin{equation}
 \Delta (\rho) = \left( \rho - \rho_{+} \right) \left( \rho -
  \rho_{-} \right),
        \;  \; \;
   R (\rho) = \rho \left( 1 - \frac{\rho_{-}}{\rho} \right)^{\alpha^{2}/(1
  +\alpha^{2})} ,  \nonumber  \\
\end{equation}
and $A_t$ is the $t$-component of the gauge potential $A_\mu$.
 The outer and `inner' horizons $\rho_{+}$ and $\rho_{-}$ are given
 by the mass $M$, the electric charge of the black hole $Q$ and
 $\alpha$ as
\begin{equation}
 \rho_{\pm} = \frac{(1+\alpha^2) (M \pm \sqrt{
  M^{2} - \left( 1-\alpha^{2} \right) Q^{2}})}{\left(1 \pm
\alpha^{2} \right)}.
  \label{eqn:parametersph}
\end{equation}
$\rho = \rho_{-}$ is the curvature singularity for $\alpha \neq 0$.
 The maximum value of the charge is $\Qmax \equiv
 \sqrt{1+\alpha^2} M$. When $|Q| = \Qmax$, $\rho_{+}$
 and $\rho_{-}$ coincide, and we call it an extreme black hole.
 However, it has to be emphasized that when $\rho_{+} = \rho_{-}$,
 a naked singularity appears at $\rho = \rho_{+}$ and the area
 of black hole vanishes for $\alpha \neq 0$, and it is therefore not
 a black hole solution\cite{GM}. \\ \hspc
The temperature $T$ of the black hole is given as
\begin{equation}
 T = \frac{1}{4 \pi \rho_{+}} \left(1-\frac{\rho_{-}}{\rho_{+}}
\right)^{(1-\alpha^2)/(1+\alpha^2)} .
\end{equation}
It possesses an interesting property\cite{GM}. When $\alpha < 1$,
$T$ in the extreme limit vanishes, whereas it diverges in the case
 of $\alpha > 1$, and has the non-zero finite value $1 / 8 \pi M$
 (as the Schwarzschild black hole) for $\alpha = 1$. \\ \hspc
Here we consider a neutral and massless scalar field which
 does not couple to the dilaton field\cite{Shiraishi2}, which is
 described by the Klein--Gordon equation
\begin{equation}
  \Phi_{,\mu}^{~;\mu} = 0  \; .
 \label{eqn:KGeq}
\end{equation}
The energy emission rate of Hawking radiation is
 given\cite{Hawking} by
\begin{equation}
 \frac{\mbox{d} M}{\mbox{d} t} = -  \frac{1}{2 \pi}
 \sum_{l,m} \int_{0}^{\infty}
\frac{\omega \: (1 - |A|^2)}{\exp \left[ \omega / \: T
\right] - 1} \: \mbox{d} \omega ,
 \label{eqn:dMsph}
\end{equation}
where $l$, $m$ are the angular momentum and its azimuthal component,
 $\omega$ is the energy of the particle, and $|A|^2$ is
 a reflection coefficient in a scattering problem
 for the scalar field $\Phi$. The Klein-Gordon equation
 (\ref{eqn:KGeq}) in this black hole spacetime
 can be made separable, by setting
\begin{equation}
 \Phi = \frac{\chi(\rho^*)}{R(\rho)} \: S(\theta) \:
\mbox{e}^{\mbox{\scriptsize i} m \varphi} \:
\mbox{e}^{- \mbox{\scriptsize i} \omega t} . \label{eqn:Phiputform}
\end{equation}
Then, Eq.(\ref{eqn:KGeq}) is reduced to the Legendre equation
 for $S(\theta)$ and the radial equation,
\begin{equation}
 \left[ \frac{\mbox{d}^2}{{\mbox{d} {\rho}^*}^{2}} +
 \omega ^{2} - {V}^2 (\rho)
 \right] \chi ({\rho}^*) = 0  ,
 \label{eqn:KGsph}
\end{equation}
where
\begin{eqnarray}
 V^2 (\rho) & \equiv &
\frac{\Delta (\rho)}{R^{2} (\rho)}
\left[
\frac{l(l+1)}{R^{2} (\rho)}  +
  \frac{1}{R (\rho)}
\frac{\mbox{d}}{\mbox{d} \rho}
\left(
{\Delta(\rho) \over {R^2(\rho)}}
\frac{\mbox{d}  R (\rho)}{\mbox{d} \rho}
  \right)
\right]  \: , \label{eqn:effpot} \\
\mbox{d} {\rho}^* & \equiv & \frac{ {R^2(\rho)}}{\Delta (\rho)} \;
\mbox{d} \rho . \label{eqn:tortdef}
\end{eqnarray}
The reflection coefficient $|A|^2$ can be calculated by solving
 the wave equation (\ref{eqn:KGsph}) numerically under
 the boundary condition
\begin{eqnarray}
 \chi & \rightarrow & \mbox{e}^{- \mbox{\scriptsize i}
 \omega \rho^*} + A \: \mbox{e}^{\mbox{\scriptsize i}
 \omega \rho^*}  ~~~~~~ \mbox{as}  ~~~ \rho^* \rightarrow \infty \: ,
 \nonumber \\
 \chi & \rightarrow & B \: \mbox{e}^{- \mbox{\scriptsize i}
 \omega \rho^*} ~~~~~~~~~~~~~~~ \mbox{as}
  ~~~ \rho^* \rightarrow - \infty \; .
\label{eqn:boundcond}
\end{eqnarray}
\hspc
The dependence of the temperature $T$ on $\alpha$ might be
 expected to imply that the behaviour
 of Hawking radiation, which is thermal and has an emission rate
 proportional to $T^4$, is drastically affected by
 the dilaton coupling, particularly for $\alpha > 1$,
 for which the temperature $T$ diverges in the extreme limit.
 However, as Holzhey and Wilczek\cite{HW} pointed out,
 since the effective potential $V$ (\ref{eqn:effpot})
 for $\alpha > 1$ grows infinitely high at the horizon
 in the extreme limit, the transmission probability
 $1 - |A|^2$ for particles to escape to infinity is suppressed.
 These two tendencies have opposite effects on
 Hawking radiation, and it is not clear whether
 or not the emission rate is actually suppressed. Here,
 we solve the wave equation (\ref{eqn:KGsph}) numerically
 to get the spectrum, and integrate Eq.(\ref{eqn:dMsph}).
 In this and subsequent calculations, we consider
 only the dominant modes with $l \leq 1$ since
 the contribution from higher angular momentum modes
 is suppressed by the centrifugal barrier. We integrate Eq.(\ref
{eqn:dMsph}) numerically to $\omega_{\mbox{\scriptsize max}}$
  ($\omega_{\mbox{\scriptsize max}} = 25 T$
 for the present non-rotating case), which is justified
 since the spectrum is suppressed at the high energy regime
 by the exponential decay in the Planck distribution. \\ \hspc
To see how the emission rate varies as the black hole
 reaches to the extreme limit, we plot the emission rate,
 normalized by mass of the black hole $M$, against $Q / \Qmax$
 for five values of the coupling constant: $\alpha = 0$,
 $0.5$, $1$, $1.5$, $2$. It is shown in Fig.1. Here, we assume
 the charge of the black hole is positive, without loss of generality.
\begin{figure}
\singlefig{10cm}{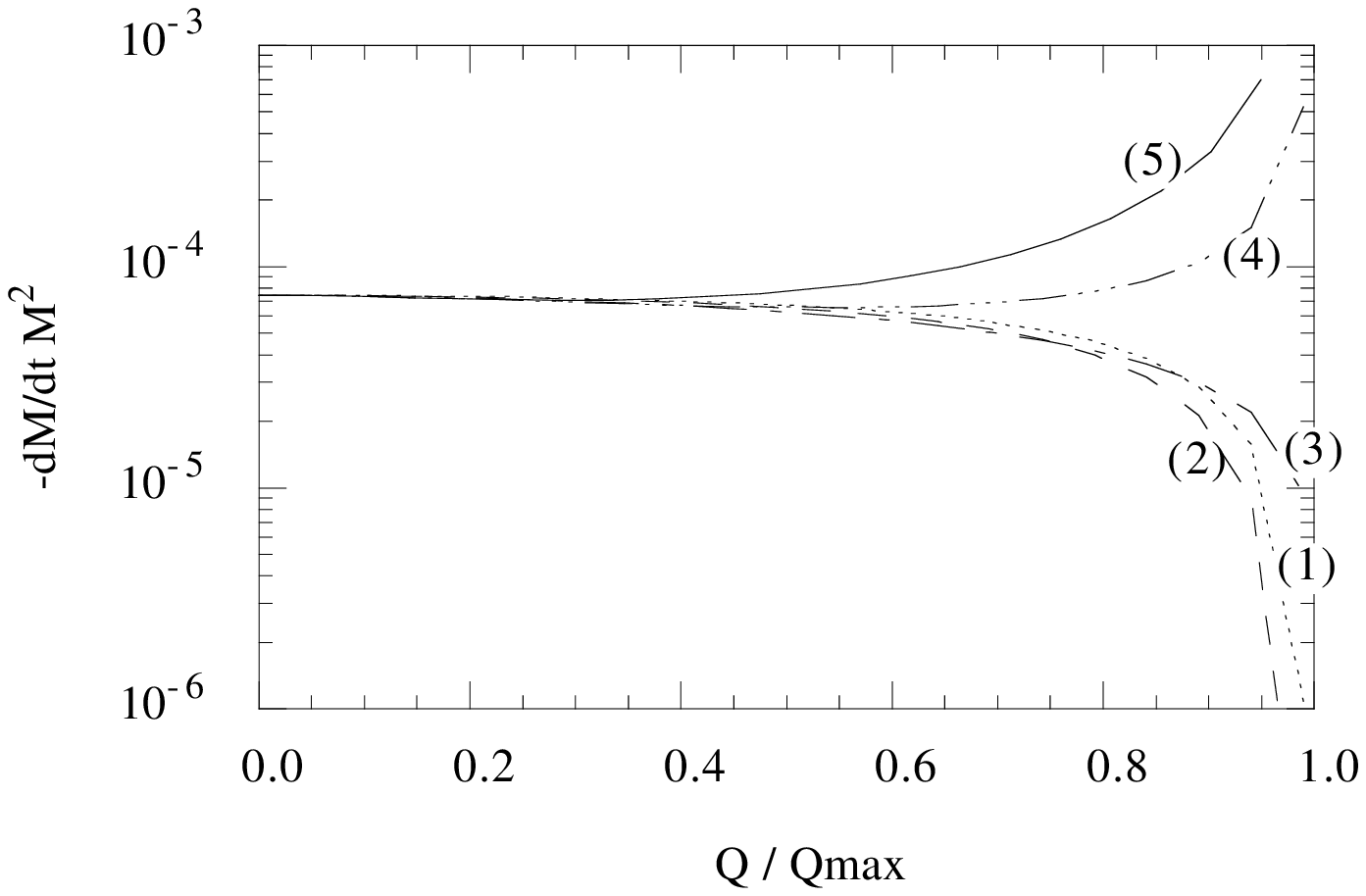}
\begin{figcaption}{fig:dMsph}{15cm}
The emission rate for the non-rotating dilatonic black holes. \\
The charge $Q$ is normalized by $\Qmax$ and the emission rate $-d M/d
t$ is normalized by $M$.  Each line corresponds to (1):$\alpha = 0$,
(2):$\alpha = 0.5$, (3):$\alpha = 1$, (4):$\alpha = 1.5$, and (5):$\alpha =
2$, respectively.
\end{figcaption}
\end{figure}
In this figure, we see that although the emission rates
 for each value of $\alpha$ coincide at $Q =0$, since
 the black hole solution, with any $\alpha$, is identically
 the Schwarzschild spacetime for $Q = 0$, the difference becomes
 large as the charge increases. In particular,
 the emission rate for $\alpha > 1$ blows up near
 the extreme limit. This means that the divergence
 of the temperature $T$ in the extreme limit overcomes
 that of the potential $V$. Furthermore, the emission rate
 (\ref{eqn:dMsph}) of the extreme black hole
 with $\alpha < 1$ is exactly zero because
 the temperature vanishes, and that for $\alpha = 1$ is
 non-zero but finite, as we see in the figure. Therefore,
 we may conclude that the behaviour of the emission
 rate in the extreme limit changes drastically
 at the value of $\alpha = 1$, as we naively expect
 from the behaviour of the temperature, despite
 the effect of the potential barrier. We may also speculate
 that nearly extreme black holes with $\alpha > 1$
 are not stable objects.

\section{Hawking Radiation from Rotating Dilatonic Black Holes}
\subsection{Rotating Dilatonic Black Holes}
\eqnum{0}
\hspc
Next, we consider Hawking radiation from rotating black holes.
 In the rotating case, we know only two exact solutions
 in the model (\ref{eqn:dilaction}): the Kerr--Newman ($\alpha = 0$)
 and the Kaluza--Klein ($\alpha = \sqrt{3}$) solution\cite{FZB}.
 Besides these two, in the $\alpha = 1$ case, an exact rotating
 black hole solution is derived by Sen\cite{Sen} in the model
 (\ref{eqn:senaction}). We first summarize these
 exact solutions and their thermodynamical properties. \\ \hspc
Firstly, the Kerr--Newman black hole solution is expressed as
\begin{eqnarray}
\mbox{d}s^2 & = & - \frac{\Delta - a^2 \sin^2\theta}{\Sigma}
 \: \mbox{d}t^2 - \frac{2 a \sin^2\theta \left( r^2 + a^2
 - \Delta \right)}{\Sigma} \: \mbox{d}t \mbox{d}\varphi
 \nonumber \\
& & + \frac{\left( r^2 + a^2 \right)^2 - \Delta
 a^2 \sin^2\theta}{\Sigma} \sin^2\theta
 \: \mbox{d}\varphi^2 + \frac{\Sigma}{\Delta}
 \, \mbox{d}r^2 + \Sigma \: \mbox{d}\theta^2 \: ,  \nonumber \\
A_t & = & \frac{Q r}{\Sigma} , \; \; \; A_\varphi
 = - \frac{a Q r \sin^2\theta}{\Sigma} \; ,
\label{eqn:KNsol}
\end{eqnarray}
where the functions $\Delta$ and $\Sigma$ are defined by
\begin{equation}
 \Delta \equiv r^2 - 2 M r + a^2 + Q^2 \; , \; \; \;
 \Sigma \equiv r^2 + a^2 \cos^2\theta \;  .
\end{equation}
The coordinate $r$ and $\rho$ in the previous section are related
 by $r = \rho - \rho_{-}$ in the spherically symmetric case.
 The temperature $T$ and the angular velocity $\OmH$ are given by
\begin{eqnarray}
T ~ & = & \frac{1}{2 \pi} \frac{\sqrt{M^2 - a^2 - Q^2}}{\rh^2 + a^2}
 \: ,  \label{eqn:TKN} \\
\OmH & = & \frac{a}{\rh^2 + a^2} \label{eqn:OmHKN} \: ,
\end{eqnarray}
where
\begin{equation}
\rh = M + \sqrt{M^2 - a^2 - Q^2}
\end{equation}
is the horizon radius, $M$, $Q$, and $J = M a$ are the mass, the charge,
 and the angular momentum of the black hole, respectively. \\
\hspc
Secondly, the Kaluza--Klein black hole solution is
 derived by a dimensional reduction of the boosted
 5-dimensional Kerr solution to four dimensions\cite{FZB,GW}.
 It is given by
\begin{eqnarray}
 \mbox{d}s^{2} & = &- \frac{\Delta -a^2 \sin^2 \theta}{B
\Sigma} \mbox{d}t^{2} - 2 a \sin^{2} \theta \frac{1}{
\sqrt{1-v^{2}}} \frac{Z}{B} \mbox{d}t \mbox{d} \varphi
  \nonumber \\
	&   & + \left[ B \left( r^{2} + a^{2} \right) + a^{2}
 \sin^{2} \theta
	  \;    \frac{Z}{B} \right] \sin^{2} \theta \; \mbox{d}
 \varphi^{2} +        \frac{B \Sigma}{\Delta} \mbox{d}r^{2}
 + B \Sigma \,
    \mbox{d} \theta^{2} \: , \nonumber  \\
 A_{t}  & = &\frac{v}{2 \left( 1-v^{2} \right) } \frac{Z}{B^{2}}, \; \; \;
 A_{ \varphi } = \; - a \sin^{2} \theta \frac{v}{2 \sqrt{1-v^{2}}}
\frac{Z}{B^{2}}, \; \; \;
 \phi = \; - \frac{\sqrt{3}}{2} \ln B \; ,
  \label{eqn:rotKKBHsol}
\end{eqnarray}
where
\begin{equation}
 \Delta \equiv r^{2} - 2 \mu r + a^{2}, \; \; \;
 \Sigma \equiv r^{2} + a^{2} \cos^{2} \theta, \; \; \;
 Z \equiv \frac{2 \mu r}{\Sigma}, \; \; \;
 B \equiv \left( 1 + \frac{v^{2} Z}{1-v^{2}} \right)^{\frac{1}{2}}   .
\label{eqn:funcdefKK}
\end{equation}
The physical mass $M$, the charge $Q$, and the angular momentum
 $J$ are expressed by the parameters $v$, $\mu$, and $a$, as
\begin{equation}
 M = \mu\left[1 +  {v^2  \over 2(1-v^2)} \right] \; ,  \; \; \;
 Q = {\mu v \over 1-v^2} \; ,  \; \; \;
 J = {\mu a \over \sqrt{1-v^2}} \; .
 \label{eqn:physKK}
\end{equation}
The horizon radius is given by
\begin{equation}
 \rh = \mu + \sqrt{ \mu^{2} - a^{2}} \; ,
\label{eqn:horizonKK}
\end{equation}
and then the regular horizon exists if
\begin{equation}
 \mu^2 \geq a^2 \; ,
 \label{eqn:horizoncond}
\end{equation}
and this condition may be rewritten as
\begin{equation}
  \left({J \over M^2}\right)^2 \leq \frac{1}{4}
\left[ 2- 10 \left({Q \over M} \right)^2 - \left({Q \over M} \right)^4
+ 2 \left(1+ 2 \left({Q \over M} \right)^2 \right)^{3/2}
\right] \; . \label{eqn:parameterrangeKK}
\end{equation}
The parameter range of the condition (\ref{eqn:parameterrangeKK})
is shown in Fig.2. It should be noted again that the solutions
 with $|Q| = \Qmax \: (= 2 M)$ are not black hole solutions and
these points are indicated by small circles in Fig.2.
\begin{figure}
\singlefig{10cm}{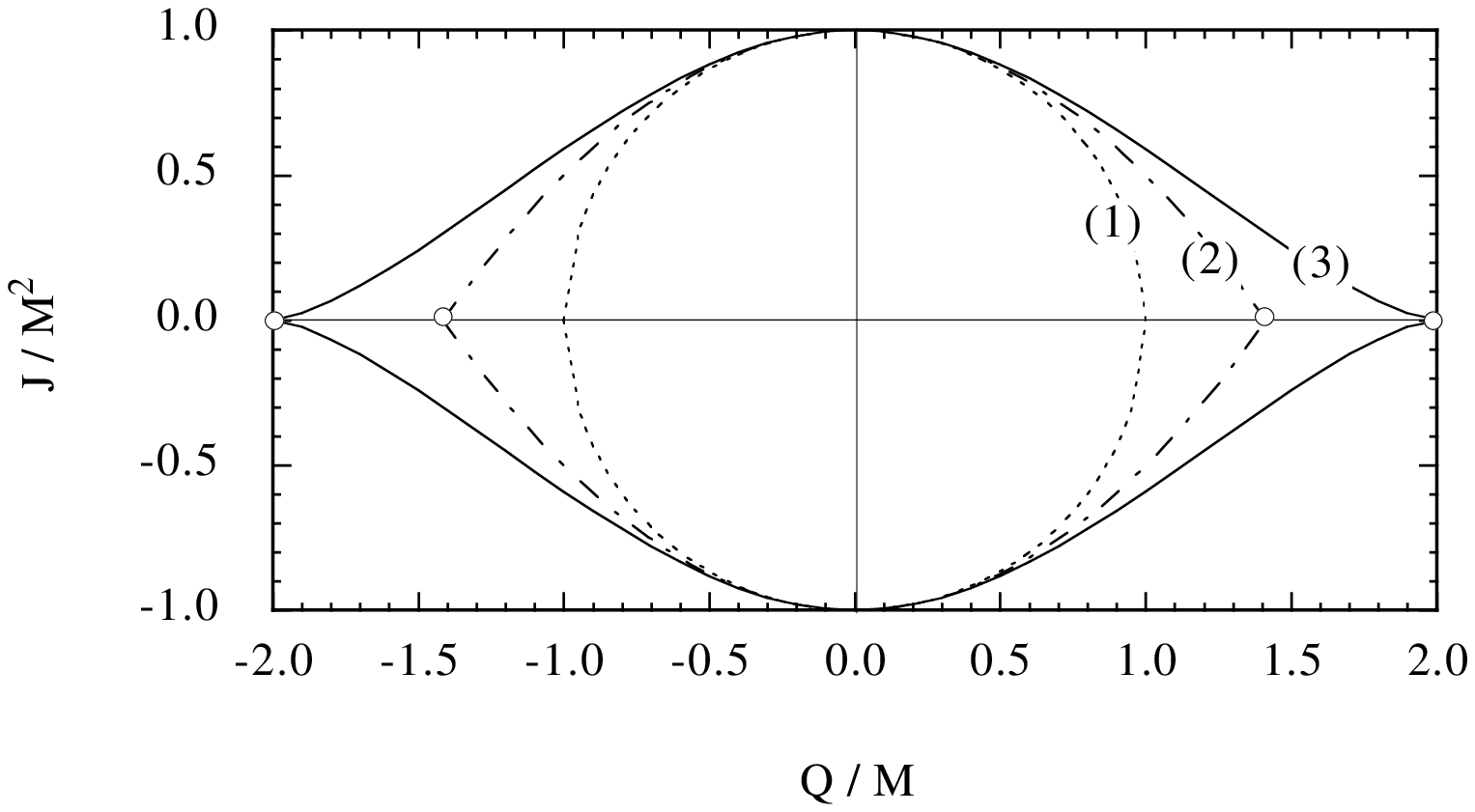}
\begin{figcaption}{fig:ParameterRange}{15cm}
The parameter ranges of three types of black hole. \\
The extreme lines are shown in $Q/M$--$J/M^2$ plane for (1):the
 Kerr--Newman, (2):the Sen, and (3):the Kaluza--Klein black holes.
 The region
inside each line guarantees a regular event horizon, except the points
denoted by a small circle where a naked singularity appears.
\end{figcaption}
\end{figure}
\\ \hspc
As for the thermodynamical properties of this black hole, we find
 that the temperature $T$ and the angular velocity $\OmH$ are
 given as
\begin{eqnarray}
T ~ & = & \frac{\sqrt{1 - v^2}}{2 \pi} \frac{\sqrt{\mu^2
- a^2}}{\rh^2 + a^2} \label{eqn:TKK} \; , \\
\OmH & =  & \frac{a \sqrt{1 - v^{2}}}{\rh^{2}
+ a^{2}} \; . \label{eqn:OmKK}
\end{eqnarray}
The temperature $T$ in the limit of $|Q| \rightarrow \Qmax$
  for the non-rotating black hole diverges, as was pointed out
 in the previous section. However, the temperature $T$ of
 the extreme rotating black hole ($\mu = |a|$) vanishes
 from Eq.(\ref{eqn:TKK}). When we take
 the limit $|Q| \rightarrow \Qmax$, keeping the black hole
 extreme with $J \neq 0$ (whereas $J \rightarrow 0$ in the limit),
 the limiting value is still zero, and different from that
 of the non-rotating case. That is, the temperature is discontinuous
 at $|Q| = \Qmax$ where a naked singularity appears.
 A similar feature is found in the behaviour of $\OmH$.
 If we take a limit $|Q| \rightarrow \Qmax$, $\OmH$ of
 a rotating black hole diverges, whereas $\OmH$ of
 a non-rotating black hole is zero. The fact that $J$ vanishes
while $\OmH$
 diverges in the limit
 $|Q| \rightarrow \Qmax$ is understood by observing
 that the area of the black hole vanishes in that limit. Those features
are illustrated in Fig.3.
\begin{figure}
\begin{center}
\segmentfig{11cm}{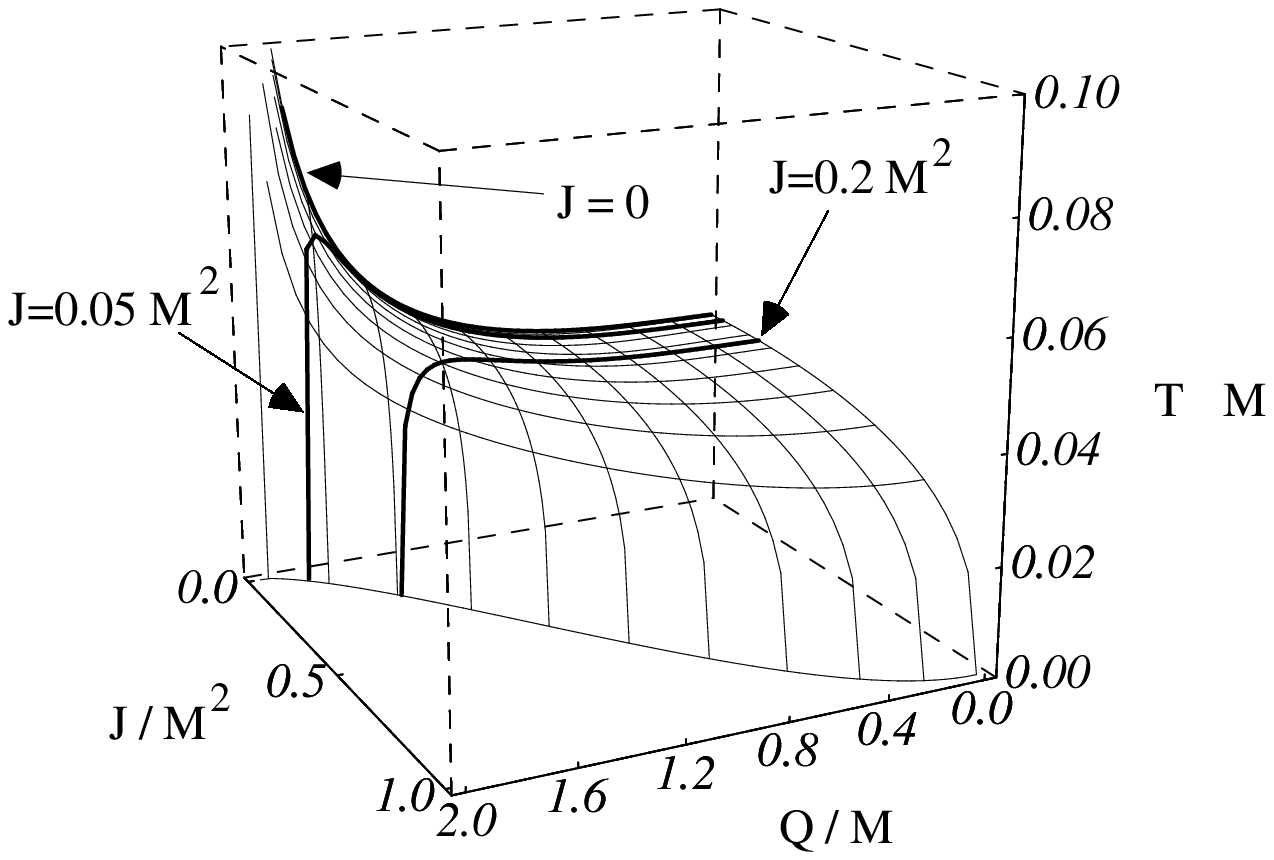}{(a)}
\vskip 1cm
\segmentfig{10cm}{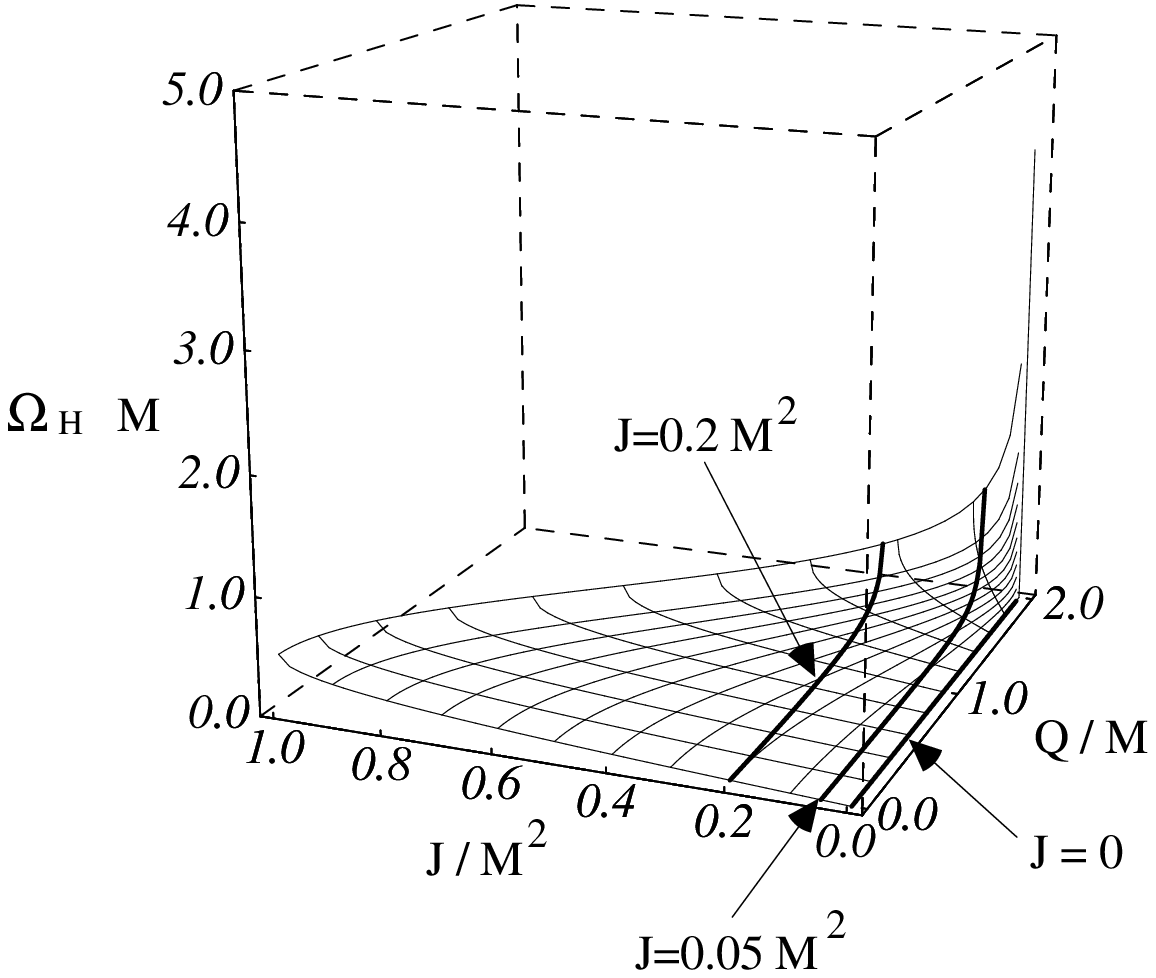}{(b)}
\end{center}
\begin{figcaption}{fig:thermKK}{15cm}
The thermodynamical behaviour of the Kaluza--Klein  black hole.\\
The behaviour of (a):the temperature $T$, and (b):the angular velocity
$\OmH$ is depicted on $Q/M$--$J/M^2$ plane ($Q \geq 0$, $J \geq 0$).
The constant angular momentum lines are drawn in solid lines for $J =
0$, $0.05 M^2$, and $0.2 M^2$.
\end{figcaption}
\end{figure}
\\ \hspc
Thirdly, the Sen black hole\cite{Sen}, which is a solution in the action
(\ref{eqn:senaction}), is expressed as
\begin{eqnarray}
\mbox{d}s^2 & = & -\frac{\Delta - a^2 \sin^2\theta}{\Sigma} \: \mbox
{d}t^2 - \frac{4 \mu r a \cosh^2\beta \sin^2\theta}{\Sigma} \: \mbox
{d}t \mbox{d} \varphi \nonumber \\
& & + \frac{\Sigma}{\Delta} \: \mbox{d} r^2 + \Sigma \: \mbox{d}
\theta^2
 + \frac{\Lambda}{\Sigma} \sin^2\theta \; \mbox{d} \varphi^2 \: ,
\nonumber \\
A_{t} & = & \frac{1}{\sqrt{2}} \frac{\mu r \sinh2\beta}{\Sigma} \: , \;
\; \; A_{\varphi} = - \frac{a}{\sqrt{2}} \sin^2\theta \; \frac{\mu r
\sinh2\beta}{\Sigma} \: , \nonumber \\
\phi & = & - \frac{1}{2} \: \ln \frac{\Sigma}{r^2 + a^2 \cos^2\theta} \: ,
\; \; \; B_{t \varphi} = 2 a \sin^2\theta \; \frac{\mu r \sinh^2\beta}
{\Sigma} \; ,
\label{eqn:Sensol}
\end{eqnarray}
where the functions $\Delta$, $\Sigma$, and $\Lambda$ are defined by
\begin{eqnarray}
 \Delta & \equiv & r^2 - 2 \mu r + a^2 , \; \; \; \Sigma \equiv r^2 + a^2
\cos^2\theta + 2 \mu r \sinh^2\beta \: , \nonumber \\
 \Lambda & \equiv & \left( r^2 + a^2 \right) \left( r^2 + a^2 \cos^2
\theta \right) + 2 \mu r a^2 \sin^2\theta \nonumber \\
& & + 4 \mu r \left( r^2 + a^2 \right) \sinh^2\beta + 4 \mu^2 r^2
\sinh^4\beta \; .
\label{eqn:Senfuncdef}
\end{eqnarray}
\hspc
The antisymmetric two rank tensor $B_{\mu\nu}$ generates the axion
field $H_{\mu\nu\rho}$, together with $A_{\mu}$, by
\begin{equation}
 H_{\mu\nu\rho} = \left( \partial_{\mu} B_{\nu\rho} - 2 A_{\mu} F_
{\nu\rho} \right) + \mbox{[cyclic permutations]} \; .
\end{equation}
The mass $M$, the charge $Q$, and the angular momentum $J$ are given
by parameters $\mu$, $\beta$, and $a$ as
\begin{equation}
 M = \frac{\mu}{2} \left( 1+\cosh2\beta \right), \; \; \; Q = \frac{\mu}
{\sqrt{2}} \sinh2\beta, \; \; \; J = \frac{a \mu}{2} \left( 1+\cosh2\beta
\right) \: ,
\end{equation}
and the horizon radius is given by the same equation as Eq.(\ref
{eqn:horizonKK}). The condition for the solution to be a black hole is
also the same as Eq.(\ref{eqn:horizoncond}), which is now rewritten as
\begin{equation}
|J| \leq M^2 - \frac{Q^2}{2} \: .
\label{eqn:parameterrangeSen}
\end{equation}
The parameter range of the condition (\ref{eqn:parameterrangeSen}) is
also shown in Fig.2. \\ \hspc
This black hole has similar thermodynamical properties to the
Kaluza--Klein solution. The temperature $T$ and the angular velocity
$\OmH$ of this black hole are given by
\begin{eqnarray}
  T & = & \frac{\sqrt{\left( 2 M^2 - Q^2 \right)^2 - 4 J^2}}{4 \pi M \left[
2 M^2 - Q^2 + \sqrt{\left( 2 M^2 - Q^2 \right)^2 - 4 J^2} \right]} \; ,
\nonumber \\
 \OmH & = & \frac{J}{M \left[ 2 M^2 - Q^2 + \sqrt{\left( 2 M^2 - Q^2
\right)^2 - 4 J^2} \right]} \; ,
\end{eqnarray}
and these quantities are discontinuous at $|Q| = \Qmax \: (= \sqrt{2}
M)$, although they never diverge but approach finite values. The
behaviour of these quantities is shown in Fig.4.
\begin{figure}
\begin{center}
\segmentfig{11cm}{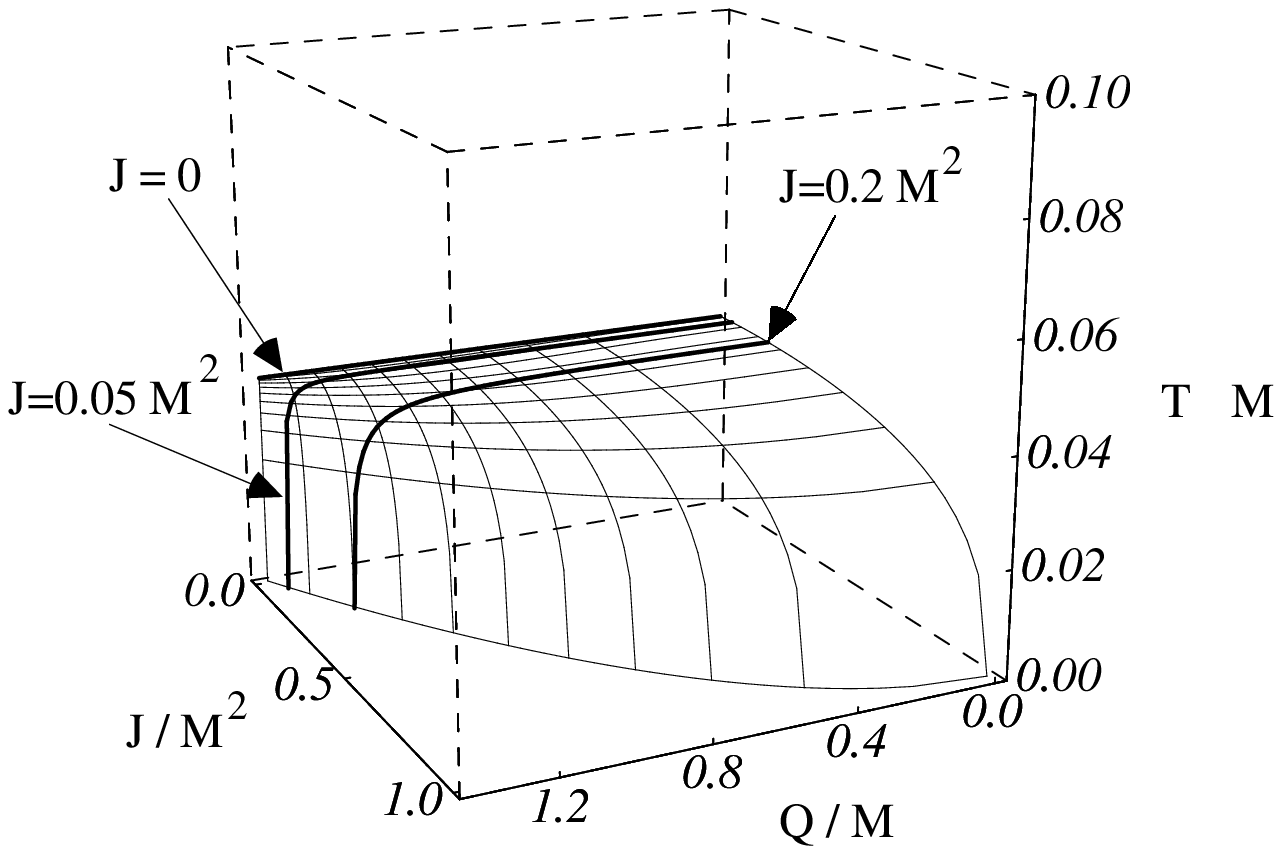}{(a)}
\vskip 1cm
\segmentfig{10cm}{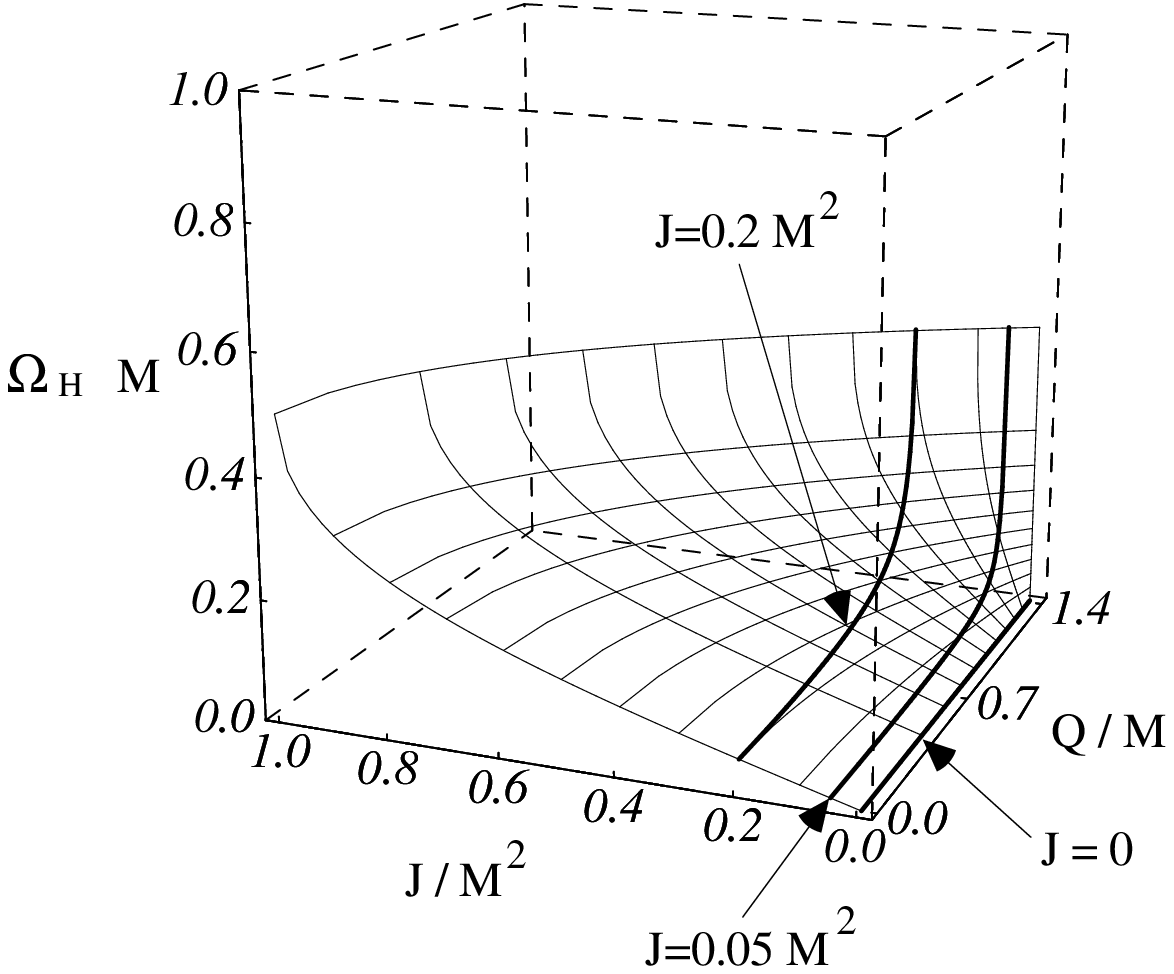}{(b)}
\end{center}
\begin{figcaption}{fig:thermSen}{15cm}
The thermodynamical behaviour of the Sen  black hole.\\
The same figures as Fig.3 are depicted for the Sen black hole.
\end{figcaption}
\end{figure}
\\ \hspc
These discontinuities indicate that the emission rate of Hawking
radiation may be completely different from that of the non-rotating case.
In addition to the thermal effect of the temperature and the effective
potential, which we considered in the previous section, new effects by
the angular velocity are important in the rotating cases: in
other words, superradiance.
\subsection{Hawking Radiation of Rotating Black Holes}
\hspc
Here, we discuss the radiation from rotating dilatonic black holes
when the black hole charge is conserved. Hereafter, we can
 assume that $Q$
and $J$ are positive without loss of generality. The Klein--Gordon
equation (\ref{eqn:KGeq}) for the neutral massless scalar field is
separated into the spheroidal equation
\begin{equation}
 \left[ \frac{1}{\sin\theta} \frac{\mbox{d}}{\mbox{d} \theta} \left(
\sin\theta \frac{\mbox{d}}{\mbox{d} \theta} \right) - \left\{ a^2
\omega^2 \sin^2\theta + \frac{m^2}{\sin^2\theta} \right\} \right] S
(\theta) = - \lambda S(\theta)
\label{eqn:Spheroidaldil}
\end{equation}
and the radial equation
\begin{equation}
 \left[ \frac{\mbox{d}^{2}}{\mbox{d} r^{* 2}}
+ \left(\omega -  m \Omega (r) \right)^2
- V^2(r)  \right]
   \chi  \left( r^* \right) = 0  ,
  \label{eqn:radialeqn}
\end{equation}
by setting
\begin{equation}
 \Phi = \frac{\chi(r^*)}{R(r)} \: S(\theta) \: \mbox{e}^{\mbox
{\scriptsize i} m \varphi} \mbox{e}^{- \mbox{\scriptsize i} \omega t} .
\end{equation}
Here the tortoise coordinate $r^*$ is defined by
\begin{equation}
 \mbox{d} r^* \equiv \frac{R^2(r)}{\Delta(r)} \mbox{d} r \; .
\end{equation}
The functions $\Omega$, $R$, and $V$ are defined
for the Kerr--Newman black hole as
\begin{displaymath}
\! \! \! \! \! \! \! \! \! \! \! \! \! \! \! \! \! \! \! \! \! \! \! \! \! \!
\! \! \! \! \! \!
\! \! \! \! \! \! \! \! \! \! \! \! \! \! \! \! \! \! \! \! \! \! \! \! \! \!
\! \! \! \! \! \!
\! \! \! \! \! \! \! \! \! \! \! \! \! \! \! \!
\Omega(r) \equiv (2 M r -Q^2) \frac{a}{R^4(r)} \: , \; \; \; \;
R^2(r) \equiv \left. \Sigma \right|_{\theta=0} = r^2 + a^2 \: ,
\end{displaymath}
\begin{equation}
V^2(r) \equiv {\Delta (r) \over R^2 (r) }
\left\{ \frac{\lambda }{R^{2}(r)}
  + \frac{1}{R(r)} \frac{\mbox{d}}{\mbox{d} r}
\left[ \frac{\Delta (r)}{R^2(r)}
\frac{\mbox{d}  R(r)}{\mbox{d}r} \right]
- {m^2 a^2 \over R^6 (r) }
(r^2 + a^2 - Q^2 + 2 M r)
\right\}  ,
\label{eqn:KGKNfncdef}
\end{equation}
for the Kaluza--Klein black hole as
\begin{displaymath}
\! \! \! \! \! \! \! \! \! \! \! \! \! \! \! \!
\Omega(r) \equiv {2 \mu r \over \sqrt{1-v^2}}{a
\over R^4(r)} \: , \; \; \;
 R^{2}(r) \equiv B \Sigma |_{\theta =0} =
(r^{2} + a^{2})  \left( 1 + \frac{v^{2}}{1-v^{2}}
\frac{2 \mu r}{r^{2} + a^{2}}
   \right)^{\frac{1}{2}} \: ,
\end{displaymath}
\begin{equation}
V^2(r) \equiv {\Delta (r) \over R^2 (r) }
\left\{ \frac{\lambda }{R^{2}(r)}
  + \frac{1}{R(r)} \frac{\mbox{d}}{\mbox{d} r}
\left[ \frac{\Delta (r)}{R^2(r)}
\frac{\mbox{d}  R(r)}{\mbox{d}r} \right]
- {m^2 a^2 \over R^6 (r) }
\left( r^2 +a^2 +{2 \mu r \over 1-v^2} \right)
\right\}
, \label{eqn:KGKKfncdef}
\end{equation}
and for the Sen black hole as
\begin{displaymath}
\! \! \! \! \! \! \! \! \! \! \! \! \! \! \! \! \! \! \! \! \! \! \! \! \! \!
\! \! \! \! \! \!
\! \! \! \! \!
\Omega(r) \equiv 2 \mu r \cosh^2 \beta \frac{a}{R^4(r)} \: , \; \; \;
R^2(r) \equiv \Sigma |_{\theta = 0} = r^2 + a^2 + 2 \mu r \sinh^2 \beta \;
,
\end{displaymath}
\begin{equation}
V^2(r) \equiv {\Delta (r) \over R^2 (r) }
\left\{ \frac{\lambda}{R^2(r)} + \frac{1}{R(r)} \frac{\mbox{d}}{\mbox
{d} r} \left[ \frac{\Delta(r)}{R^2(r)} \frac{\mbox{d} R(r)}{\mbox{d} r}
\right]
- \frac{m^2 a^2}{R^6(r)} \left( r^2 + a^2 + 2 \mu r \cosh 2\beta \right)
\right\}
 . \label{eqn:KGSenfncdef}
\end{equation}
\hspc
The emission rates of energy and angular momentum\cite{Hawking} are
given by
\begin{eqnarray}
 \frac{\mbox{d} M}{\mbox{d} t} & = & - \frac{1}{2 \pi} \sum_{l, m}
\int^{\infty}_{0}
          \frac{\omega \left( 1 - | A |^{2} \right)}{\exp \left[ \left( \omega
- m \OmH \right) / \: T \right] - 1} \: \mbox{d} \omega \; , \nonumber
\\
 \frac{\mbox{d} J}{\mbox{d} t} & = & - \frac{1}{2 \pi} \sum_{l, m}  \int^
{\infty}_{0}
          \frac{m  \left( 1 - | A |^{2} \right)}{\exp \left[ \left( \omega - m
\OmH \right) / \: T \right] - 1} \: \mbox{d} \omega  \; .
    \label{eqn:radrot}
\end{eqnarray}
The reflection coefficient $|A|^2$ is calculated by solving the wave
equation (\ref{eqn:radialeqn}) under the boundary condition
\begin{eqnarray}
 \chi & \rightarrow & \mbox{e}^{- \mbox{\scriptsize i} \omega r^*} + A
\: \mbox{e}^{\mbox{\scriptsize i} \omega r^*}  ~~~~~~ \mbox{as}  ~~~
r^* \rightarrow \infty \; , \nonumber \\
 \chi & \rightarrow & B \: \mbox{e}^{- \mbox{\scriptsize i} \widetilde
{\omega} r^*} ~~~~~~~~~~~~~~~ \mbox{as}  ~~~ r^* \rightarrow -
\infty \; ,
\end{eqnarray}
where $\widetilde{\omega} \equiv \omega - m \OmH$. We integrate
 Eq.(\ref{eqn:radrot}) by setting the upper bound of the integration
 as
 $\omega_{\mbox{\scriptsize max}} = \max(25T, 1.5\OmH )$
 for a rotating black
hole. The eigenvalue $\lambda$ of the spheroidal equation (\ref
{eqn:Spheroidaldil}) is calculated perturbatively\cite{Bouwkamp} as
\addtolength{\abovedisplayskip}{0.2cm}
\addtolength{\belowdisplayskip}{0.2cm}
\addtolength{\jot}{0.2cm}
\begin{eqnarray}
\addtolength{\abovedisplayskip}{-0.2cm}
\addtolength{\belowdisplayskip}{-0.2cm}
\addtolength{\jot}{-0.2cm}
 \lambda & = & l(l+1) + \frac{1}{2} \left[ \frac{(2m-1) (2m+1)}{(2l-1)
(2l+3)}+1 \right] a^2 \omega^2 + \frac{1}{2} \left[ \frac{(l-m-1) (l-m)
(l+m-1) (l+m)}{(2l-3) (2l-1)^3 (2l+1)} \right. \nonumber \\
& & \left. -\frac{(l-m+1) (l-m+2) (l+m+1) (l+m+2)}{(2l+1) (2l+3)^3 (2l+
5)} \right] a^4 \omega^4 + {\cal O} (a^6 \omega^6)  .
\end{eqnarray}
This approximation is valid, since $a \omega < 1$ and the coefficient of
each term is small for all the cases we analyzed, although $\omega M > 1$
in some instances. \\ \hspc
As we mentioned before, the coupling constant dependence of the
temperature or the angular velocity is remarkable in highly charged
black holes. Hence we analyze such cases. If the black hole has a large
charge, it carries only a small angular momentum, as is seen from Fig.2.
Because of this, we consider the black holes with a small angular
momentum, which is fixed at $J = 0.01 M^2$, and vary the charge to see
how the emission rates for each solution change in the extreme limit.
The result is shown in Fig.5. The charge is normalized by the maximal
value  $\Qex$  for the black hole with $J = 0.01 M^2$, which is a little
less than $\Qmax$ (See Fig.2). The values of $\Qex$ are $0.999 \:
\Qmax$, $0.995 \: \Qmax$, and $0.972 \: \Qmax$ for the Kerr--Newman,
the Sen and the Kaluza--Klein black holes, respectively.
\begin{figure}
\begin{center}
\segmentfig{10cm}{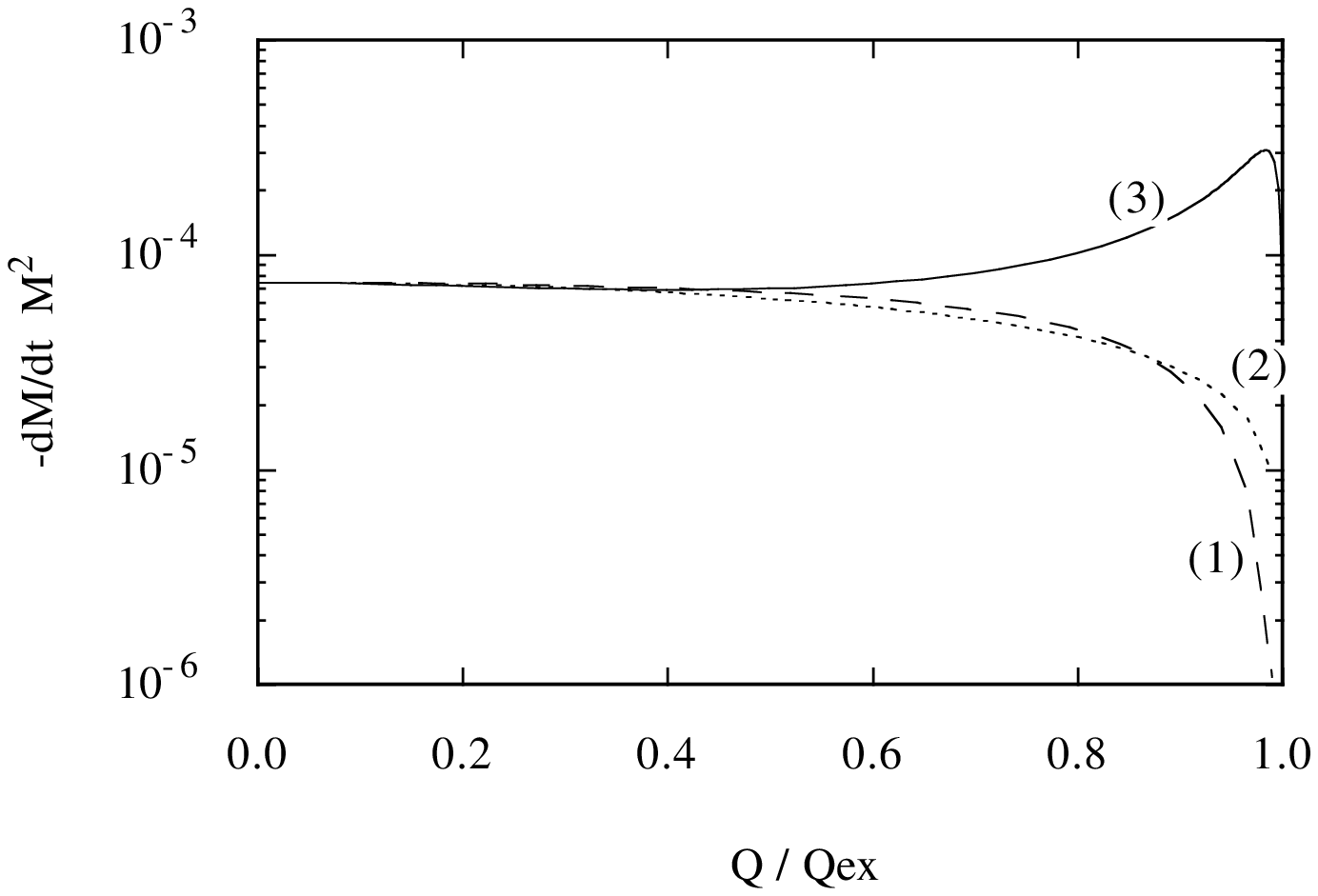}{(a)}
\vskip 1cm
\segmentfig{10cm}{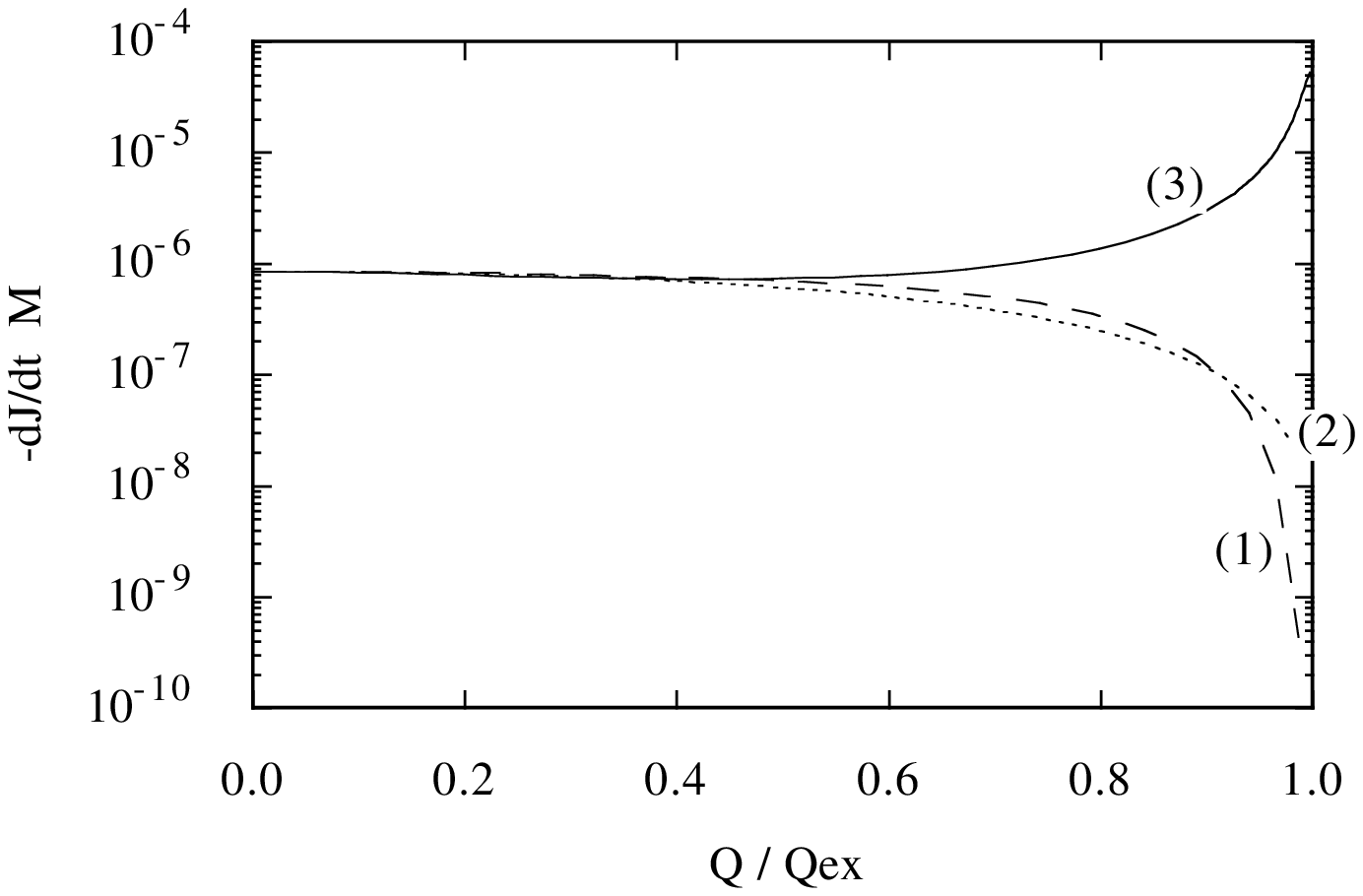}{(b)}
\end{center}
\begin{figcaption}{fig:radrot}{15cm}
The emission rate of (a):the energy $-dM/dt$ and (b):the angular
momentum $-dJ/dt$ for three types of rotating black holes. \\
$Q$ is normalized by $\Qex$. Each line corresponds to (1):the
 Kerr--Newman, (2):the Sen, and (3):the Kaluza--Klein black holes,
respectively. $\Qex$ is $0.999 \: \Qmax$, $0.995 \: \Qmax$, and $0.972
\: \Qmax$ for Kerr--Newman, for the Sen, and for the Kaluza--Klein
black hole, respectively.
\end{figcaption}
\end{figure}
In Fig.5, we see the Kaluza--Klein black hole radiates much more energy and
angular momentum near the extreme limit than
the Kerr--Newman and the Sen black holes. The behaviour of the
energy emission rates (Fig.5 (a)) is very similar to that of
 non-rotating black holes, except in the vicinity of the extreme
 limit. This is
because we have chosen a very small value for the angular momentum.
However, in the  Kaluza--Klein black hole, the emission rate drops a
little near the extreme limit and does not diverge, so we find a
different result from the non-rotating case. This is because the
temperature of the rotating Kaluza--Klein black hole vanishes in the
extreme limit whereas that of the non-rotating case is divergent in the
same limit. \\ \hspc
There appears to be a critical value of the dilaton coupling constant at
$\alpha \sim 1$, although we cannot give a definite critical value from
our analysis of exact black hole solutions. However there is another
way to investigate such a critical value in the extreme limit. The
temperatures of rotating black holes vanish in the extreme limit, and it
is known that Hawking radiation becomes purely superradiant\cite
{Unruh}, that is, the emission rates (\ref{eqn:radrot}) are
\begin{eqnarray}
 \frac{\mbox{d} M}{\mbox{d} t} & = & - \frac{1}{2 \pi} \sum_{l, m}
\int^{m \OmH}_{0}
          \omega \left( | A |^{2} -1 \right) \mbox{d} \omega
 \; , \nonumber \\
 \frac{\mbox{d} J}{\mbox{d} t} & = &- \frac{1}{2 \pi} \sum_{l, m}  \int^
{m \OmH}_{0}
          m  \left( | A |^{2} -1 \right) \mbox{d} \omega   .
    \label{eqn:radrotsup}
\end{eqnarray}
\hspc
In the previous paper\cite{KM}, we analyzed superradiance of the
rotating dilatonic black holes, which we will briefly summarize.
 To see
how superradiance depends on the dilaton coupling constant, we
considered the slowly rotating approximate solution with arbitrary
coupling constant\cite{HH,Shiraishi1}, which is given by adding an
angular momentum perturbation to the spherically symmetric solution
(\ref{eqn:nonrotsol}), as well as the three exact solutions. This solution
is expressed, in the same coordinates as the spherically symmetric
solution in the section 2, as
\begin{eqnarray}
\mbox{d}s^{2} & = & -{\Delta(\rho) \over {R^2(\rho)}}
\mbox{d}t^{2} + \frac{{R^2(\rho)}}{\Delta(\rho)} \mbox{d}\rho^{2} + {
{R}^{2}(\rho)} (\mbox{d} \theta^{2} + \sin^{2} \theta \mbox{d} \varphi^
{2} ) - 2 a f(\rho) \sin^{2}\theta \: \mbox{d} t \mbox{d} \varphi
\nonumber \\
A_{t} & = & \frac{Q}{\rho}, \; \; \; A_{\varphi} = - a \sin^{2} \theta
\frac{Q}{\rho}, \; \; \; \phi = \frac{\alpha}{1+\alpha^{2}} \ln \left(1 -
\frac{\rho_{-}}{\rho} \right) \; ,   \label{eqn:slowsol}
\end{eqnarray}
where
\begin{eqnarray}
 \Delta (\rho) & \equiv & \left( \rho - \rho_{+} \right) \left( \rho -
  \rho_{-} \right),
        \;  \; \;
   R (\rho) \equiv \rho \left( 1 - \frac{\rho_{-}}{\rho} \right)^{\alpha^
{2}/(1  +\alpha^{2})}   \; , \nonumber  \\
 f (\rho) & \equiv & \frac{\left( 1+\alpha^{2} \right)^{2}}{\left( 1-
\alpha^{2}
  \right) \left( 1-3\alpha^{2} \right) } \left({\rho \over \rho_{-}}
\right)^2
\left( 1 -   \frac{\rho_{-}}{\rho}
\right)^{2\alpha^{2}/(1  +\alpha^{2})} - \left( 1 -
  \frac{\rho_{-}}{\rho} \right)^{(1-\alpha^{2})/(1+\alpha^{2})}
      \nonumber  \\
  &  \times & \left( 1 +  \frac{\left( 1+\alpha^{2} \right)^{2}}{\left(
  1-\alpha^{2} \right) \left( 1-3\alpha^{2} \right) }
 \left({\rho \over \rho_{-}}\right)^2
  + \frac{1+\alpha^{2}}{1-\alpha^{2}} \left({\rho \over \rho_{-}}\right)
-   \frac{\rho_{+}}{\rho} \right) \; ,
\end{eqnarray}
and
\begin{equation}
 \rho_{\pm} = \frac{(1+\alpha^2) (M \pm \sqrt{
  M^{2} - \left( 1-\alpha^{2} \right) Q^{2}})}{\left(1 \pm \alpha^{2}
\right)}, \; \;    \;
 a = \frac{2 (1 + \alpha^2) J}{ (1 + \alpha^2) \rho_{+} +
(1-\alpha^{2}/3) \rho_{-}}   .
  \label{eqn:parameteraprx}
\end{equation}
This solution is valid only when the parameter $a$ is sufficiently
small.  Although
$ f(\rho) $ seems to diverge at $ \alpha = 1/\sqrt{3} $, $ \alpha = 1 $ or
$ \rho_{-} = 0 $, $f(\rho)$ approaches a finite limiting value when we
expand this function around each point. \\ \hspc
The Klein-Gordon equation is now separated into the Legendre equation
and the radial equation
\begin{equation}
   \left[ \frac{\mbox{d}^2}{{\mbox{d} {\rho}^*}^{2}} + \left( \omega -  m
\Omega (\rho) \right)^{2} - {V}^2 (\rho)
 \right] \chi ({\rho}^*) = 0 \; ,
 \label{eqn:radialslow}
\end{equation}
where
\begin{eqnarray}
 \Omega (\rho) & \equiv &
   \frac{a  f (\rho)}{ R^{2} (\rho)},\\
V^2 (\rho) & \equiv &
\frac{\Delta (\rho)}{R^{2} (\rho)}
\left[
\frac{l(l+1)}{R^{2} (\rho)}  +
  \frac{1}{R (\rho)}
\frac{\mbox{d}}{\mbox{d} \rho}
\left(
{\Delta(\rho) \over {R^2(\rho)}}
\frac{\mbox{d}  R (\rho)}{\mbox{d} \rho}
  \right)
\right],
\end{eqnarray}
and
${\rho}^*$ is defined by
\begin{equation}
 \mbox{d} {\rho}^* \equiv \frac{ {R^2(\rho)}}{\Delta (\rho)} \: \mbox{d}
\rho  \; . \label{eqn:tortoiseaprx}
\end{equation}
The emission rates by superradiance for this approximate black hole
solution are shown in Fig.6, with the calculations using the three exact
rotating black hole solutions.
\begin{figure}
\begin{center}
\segmentfig{10cm}{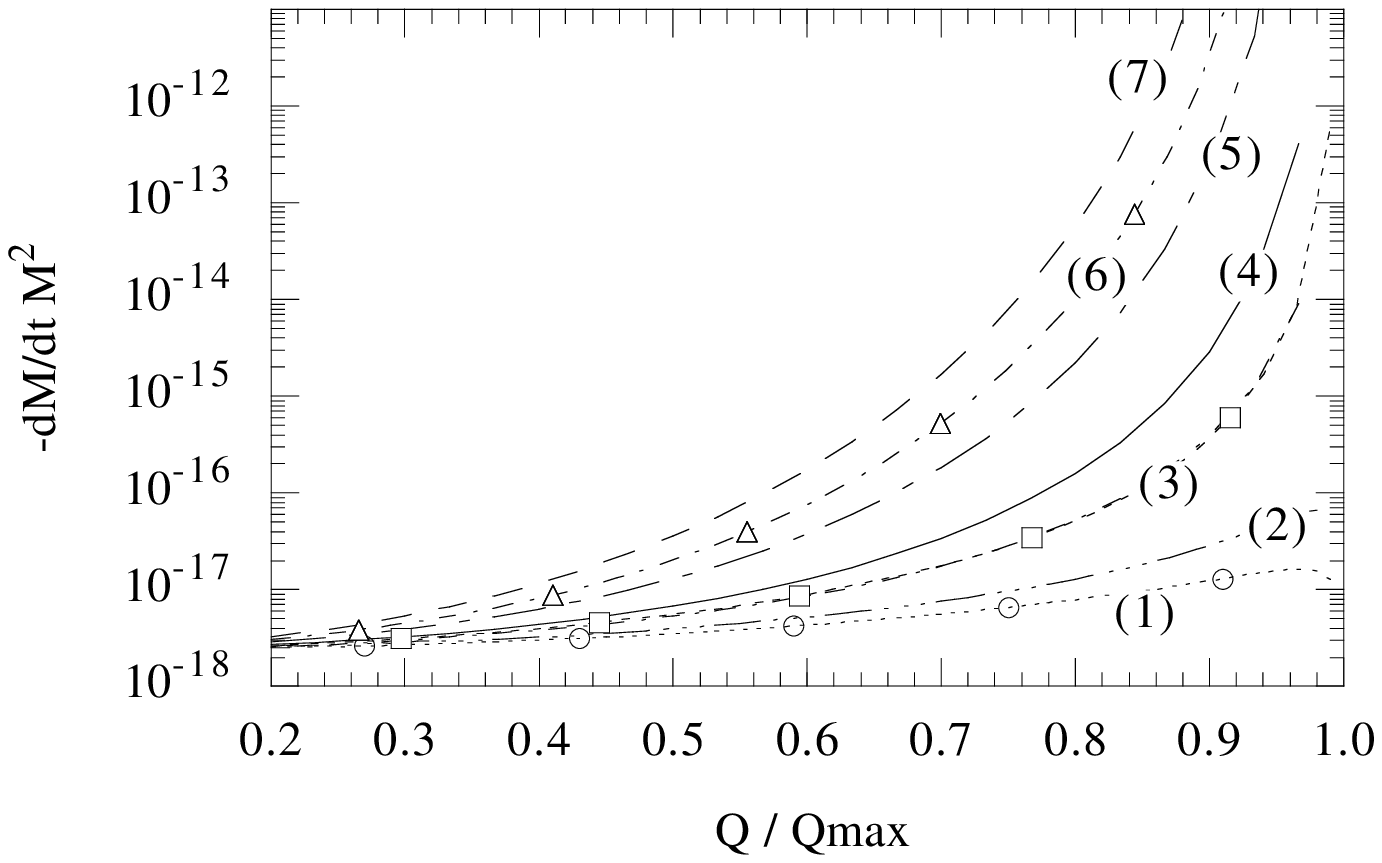}{(a)}
\vskip 1cm
\segmentfig{10cm}{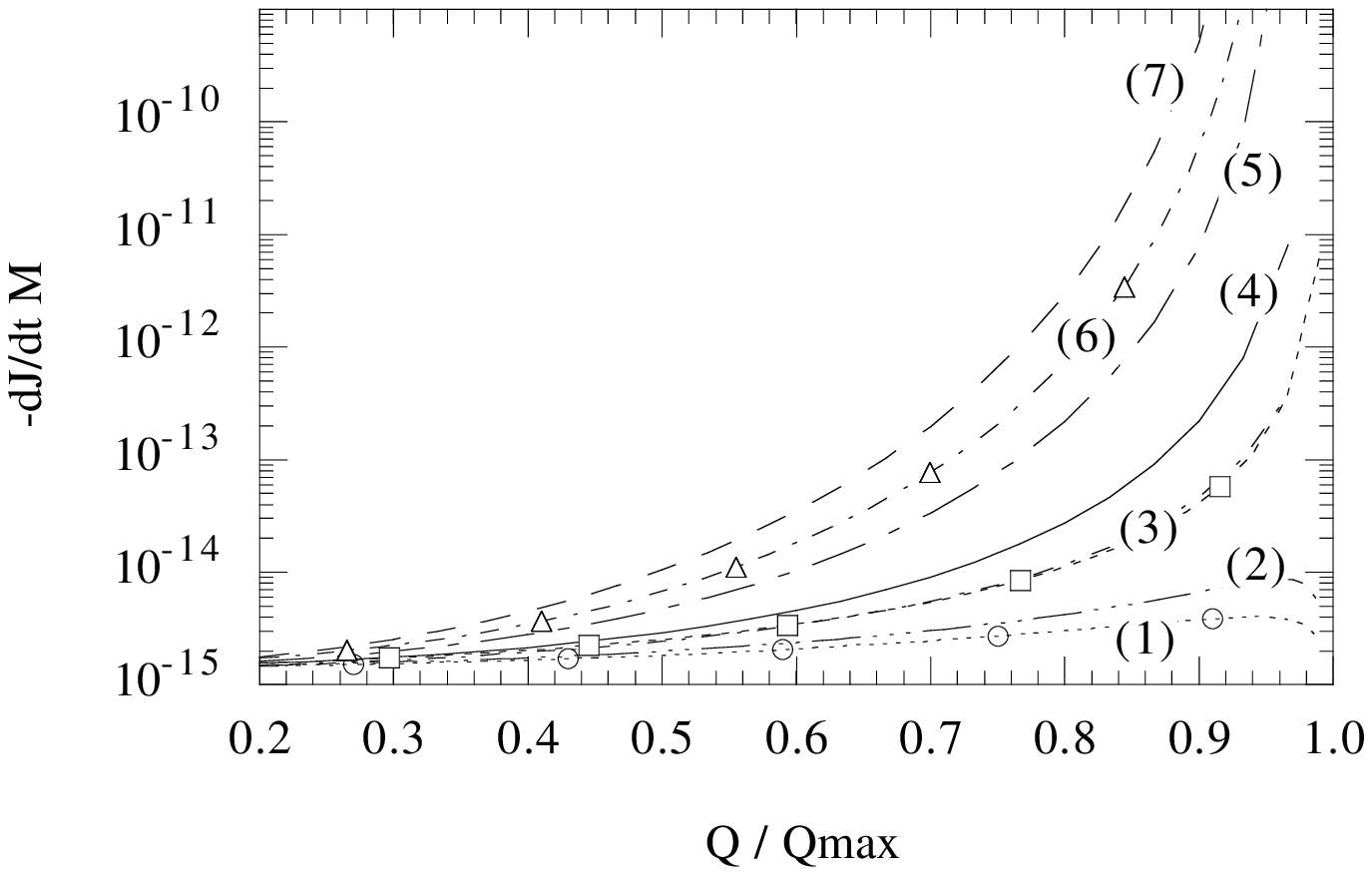}{(b)}
\end{center}
\begin{figcaption}{fig:radsup}{15cm}
Superradiance from slowly rotating black holes. \\
(a) and (b) show the energy emission rate $-dM/dt$, and the angular
momentum emission rate $-dJ/dt$, respectively. Each line corresponds
to (1):$\alpha = 0$, (2):$\alpha = 0.5$, (3):$\alpha = 0.9$, (4):$\alpha =
1.1$, (5):$\alpha = 1.5$, (6):$\alpha = \sqrt{3}$, and (7):$ \alpha = 2$.  In
addition, we plot the results for three exact solutions (the circles for
the Kerr--Newman, the squares for the Sen, and the triangles for the
Kaluza--Klein black holes). The charge is normalized by $\Qmax$.
\end{figcaption}
\end{figure}
We find that the emission rate from the large coupling constant
black holes blows up as the black hole approaches the extreme one,
whereas with small $\alpha$, the emission rate remains quite small.
 These two types of
behaviour are divided by a value of the coupling constant of about
 unity.  \\ \hspc
Again we cannot determine the exact value of the critical coupling
constant, for the following reason. As we mentioned before, this
approximate black hole solution is valid only when the angular
momentum is sufficiently small. In addition to this condition,
 there is another
requirement that must be satisfied, namely, that the black hole
charge should not be so large. This is found by observing that the
maximally charged black hole ($Q = \Qmax$) in the approximate
solution can carry an angular momentum, while the exact solution
cannot (e.g., consider the Kerr--Newman black hole). Quantitatively, the
angular velocity $\OmH \equiv \Omega(\rho_{+})$ of the black hole is
divergent in the extreme limit for $\alpha \geq 1/\sqrt{3}$ and
vanishes for smaller coupling constants, but this critical value is derived
from the approximate solution and may differ from the value of the exact
 solution, which we do not know. In fact, the angular velocity in the
extreme limit of the Sen black hole, which is a solution of the model
(\ref{eqn:senaction}) ($\alpha = 1$), is finite and non-zero, although
this solution is obtained from a different action from
 the Kerr--Newman and the
Kaluza--Klein black holes. The qualitative behaviour of the angular
velocity in the approximate solution seems to follow that of the
exact solution closely, although that of the temperature does not.
 Hence we may
give a qualitative discussion of superradiance by using this
approximate solution. \\ \hspc
We may conclude from the above that the critical coupling constant at
which the behaviour of the superradiant emission changes exists and is
about unity.  As we have already shown, the behaviour of the emission
rate by thermal radiation from the non-rotating black hole also
changes at $\alpha = 1$. Naively speaking, Hawking radiation for the
rotating black hole consists of two components, that is, thermal
radiation and superradiance. So we naturally expect that the
 emission of Hawking radiation from rotating black
holes is drastically changed at $\alpha \sim 1$.
\subsection{The Fate of Dilatonic Black Holes}
\hspc
The dependence of the emission on the coupling constant leads to a
difference in the evolution of black holes by evaporation. To
investigate the evolution of the three exact rotating solutions above, we
describe the black hole state by a pair of quantities $(Q/M, J/M^2)$, and
analyze their time variations,  which are given by the
emission rates as
\begin{eqnarray}
 \frac{\mbox{d}}{\mbox{d} t} \left( \frac{Q}{M} \right) & = & - \frac{Q}
{M^2} \frac{\mbox{d} M}{\mbox{d} t} \; , \label{eqn:evolQ} \\
\frac{\mbox{d}}{\mbox{d} t} \left( \frac{J}{M^2} \right) & = & \frac{1}
{M^2} \frac{\mbox{d} J}{\mbox{d} t} - 2 \frac{J}{M^3} \frac{\mbox{d} M}
{\mbox{d} t} \; , \label{eqn:evolJ}
\end{eqnarray}
for the three exact black hole solutions. We recognize these two
 quantities
as a vector field on the $Q/M$--$J/M^2$ plane shown in Fig.2, and show
it in Fig.7.
\begin{figure}
\begin{center}
\segmentfig{5.7cm}{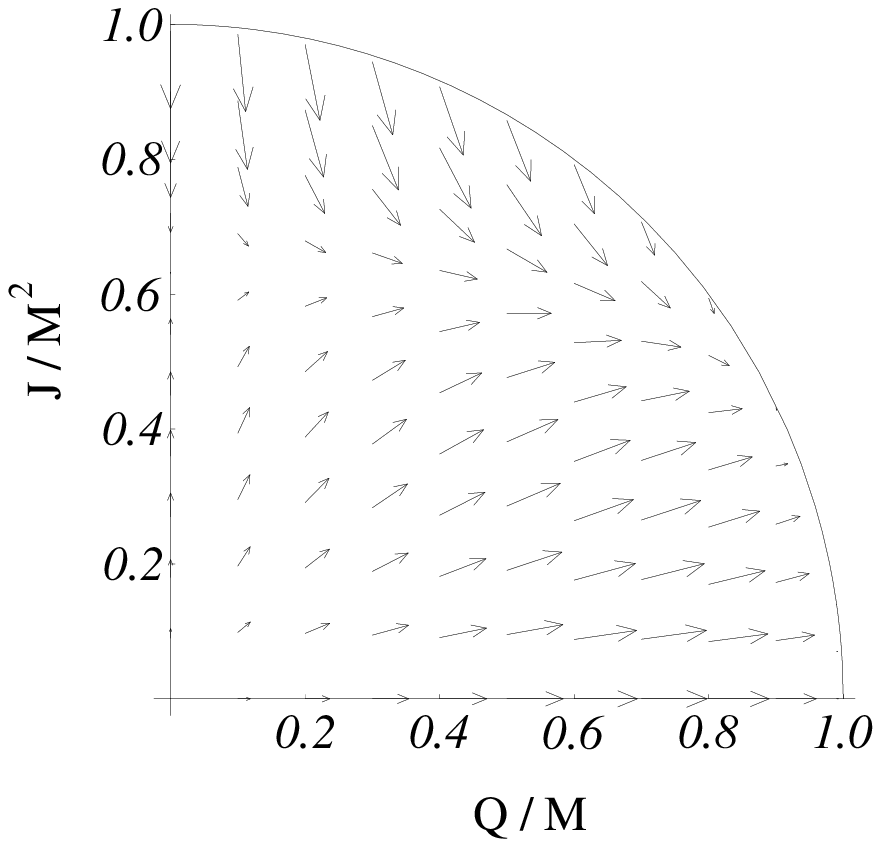}{(a)}
\hspace{2cm}
\segmentfig{7.66cm}{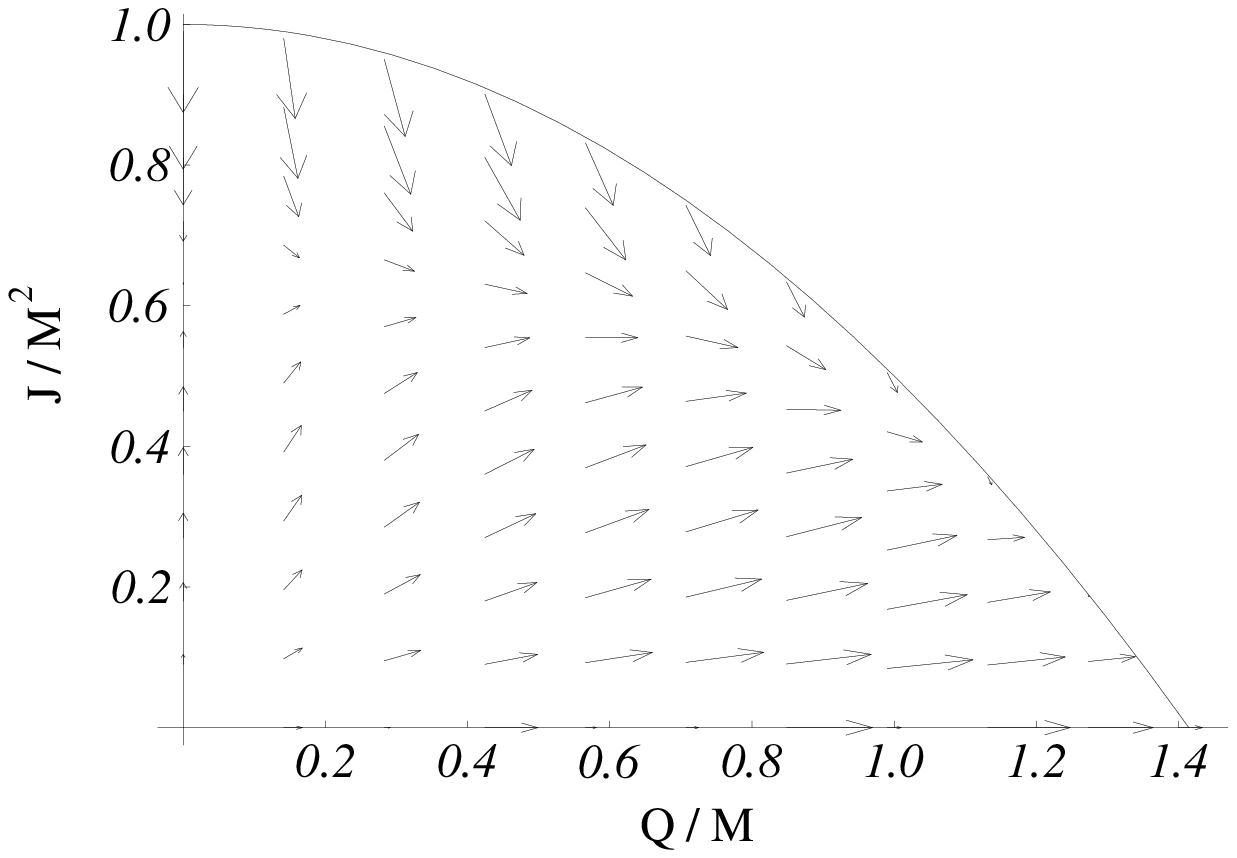}{(b)}
\end{center}
\vskip 2cm
\begin{center}
\segmentfig{16cm}{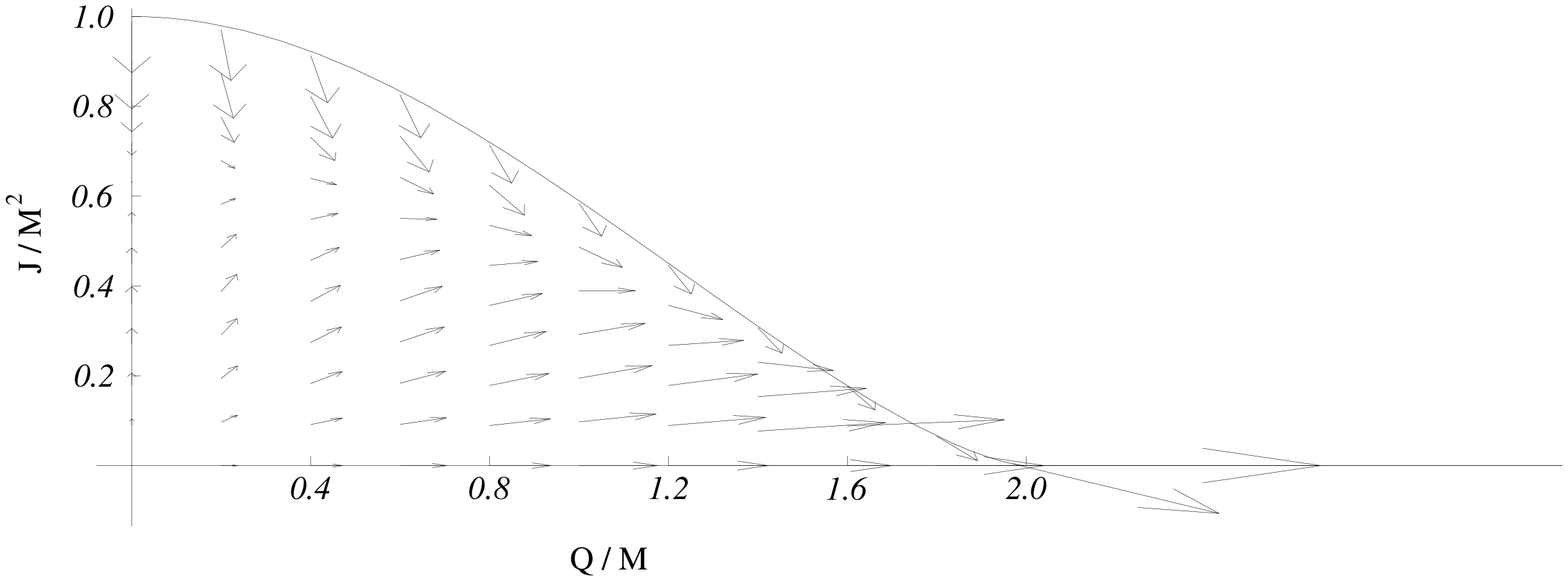}{(c)}
\end{center}
\begin{figcaption}{fig:EvolVec}{15cm}
The  evolution of three types of black hole. \\
Each figure represents
(a):the Kerr--Newman, (b):the Sen, and (c):the Kaluza--Klein black hole.
The arrow shows the direction and magnitude of
 the evolution of the black hole
by Hawking evaporation at each point.   The scale of the arrow is
enlarged 2500 times. Although the arrows
 near $Q = \Qmax$ are very small for the Kerr--Newman and
the Sen black holes,  those in the Kaluza--Klein black hole
  are considerably larger.
\end{figcaption}
\end{figure}
Since we assume that the black hole charge is conserved, and
 the black
holes lose mass energy, Eq.(\ref{eqn:evolQ}) is always positive, so $Q/
M$ increases and the black hole approaches the extreme state. In
Fig.7, near the extreme lines, each vector points to a direction inside the
extreme line, so the black hole does not evolve beyond the extreme line
and eventually approaches the $Q = \Qmax$ state. From the figure,
we can see that the Kerr--Newman black hole stops its evolution as it
approaches $Q = \Qmax$ whereas the evolution of the Kaluza--Klein
black hole is accelerated as $Q / M$ increases, and in particular,
 the evolution is very fast near $Q = \Qmax$. This is because
the emission rates of the Kaluza--Klein black hole near $Q = \Qmax$
state are very large. As we mentioned before, the $Q = \Qmax$ state is
not a black hole solution and a naked singularity appears at this point.
So it is indicated from our analysis that the Kaluza--Klein black hole
evolves rapidly into a naked singularity. As we
have already seen, the area of the Kaluza--Klein black hole vanishes as
$Q \rightarrow \Qmax$. This situation is quite similar to the
evaporation of the Schwarzschild black hole, for which the area of the
black hole vanishes and the emission rate increases infinitely large in
the final stage, where a naked singularity might appear. The Sen black
hole shows an intermediate behaviour between that of the Kerr--
Newman and the the Kaluza--Klein black holes.

\section{Discharge of Dilatonic Black Holes by Superradiance}
\eqnum{0}
\hspc
So far, we have considered only the case where the charge of the black
hole is conserved. Usually, however, black holes may create charged
particles and lose their charge. In this section we study the discharge
process by superradiance of a charged scalar field described by the
equation of motion
\begin{equation}
 \left[ \left( \nabla^\mu + \mbox{i} e A^\mu \right) \left( \nabla_\mu +
\mbox{i} e A_\mu \right)- \mu^2 \right] \Phi = 0 \; ,
 \label{eqn:KGq}
\end{equation}
where $e$ and $\mu$ are the charge and the rest mass of the particle,
respectively. Shiraishi\cite{Shiraishi2} analyzed superradiance of a
charged scalar field $\Phi$ coupled to the dilaton $\phi$ in the spherically
symmetric dilatonic black hole. Here we do not consider such a coupling
because we are only interested in the pure quantum properties
 of the dilatonic
black hole, but not the extra effects on the quantum radiation, which
come from a direct coupling between $\Phi$ and the dilaton field. \\ \hspc
The timescales of the loss of energy, angular momentum, and charge
 depend on the temperature $T$, the angular velocity $\OmH$,
and the electric potential $\PhiH$ in the Planck distribution of
 Hawking
radiation as
\begin{equation}
 \frac{1}{\exp\left[\left(\omega - m \OmH - e \PhiH\right) / T \right]}
\; .
\end{equation}
If the electric potential is large enough compared with the temperature
and the angular velocity, the dominant component of the emission is
that of the superradiant discharge process. In order to estimate
 how important the
discharge process is in Hawking radiation, we calculate
the superradiant emission rates in a spherically symmetric dilatonic
black hole, in which the electric potential $\PhiH$ is
\begin{equation}
\PhiH = \frac{Q}{\rho_{+}} \; .
\label{eqn:phiHdil}
\end{equation}
The horizon radius $\rho_{+}$ is given by Eq.(\ref{eqn:parametersph}). If
superradiance is large compared to the emission calculated in the
previous sections, where we assumed that the charge is conserved, the
discharge process is important and should not be ignored, while, if it is
small, the discharge process is not essential in Hawking radiation.
\\ \hspc
The emission rates are
\begin{eqnarray}
 \frac{\mbox{d} M}{\mbox{d} t} & = & - \frac{1}{2 \pi} \sum_{l , m , e}
\int_{\mu}^{e \PhiH} \omega \left( |A|^2 - 1 \right) \mbox{d}\omega
\label{eqn:dMRNsuper} \; , \\
 \frac{\mbox{d} Q}{\mbox{d} t} & = & - \frac{1}{2 \pi} \sum_{l , m , e}
\int_{\mu}^{e \PhiH} e \left( |A|^2 - 1 \right) \mbox{d}\omega \: , \label
{eqn:dQRNsuper}
\end{eqnarray}
where the reflection coefficient $|A|^2$ is obtained by solving the
radial wave equation
\begin{equation}
 \left[ \frac{\mbox{d}^2}{{\mbox{d} {\rho}^*}^{2}} + \left( \omega - e
\frac{Q}{\rho} \right)^{2} - \mu^2 \frac{\Delta(\rho)}{R^2(\rho)} - {V}^2
(\rho)
 \right] \chi ({\rho}^*) = 0  \; ,
 \label{eqn:Radialdis}
\end{equation}
which is derived by setting in the same way as Eq.(\ref
{eqn:Phiputform}), under the boundary condition of
\begin{eqnarray}
 \chi & \rightarrow & \mbox{e}^{- \mbox{\scriptsize i} \omega \rho^*}
+ A \: \mbox{e}^{\mbox{\scriptsize i} \omega \rho^*}  ~~~~~~ \mbox
{as}  ~~~ \rho^* \rightarrow \infty \; , \nonumber \\
 \chi & \rightarrow & B \: \mbox{e}^{- \mbox{\scriptsize i} \widetilde
{\omega} \rho^*}  ~~~~~~~~~~~~~~~ \mbox{as}  ~~~ \rho^*
\rightarrow - \infty \; ,
\end{eqnarray}
where $\widetilde{\omega}$ is now defined as $\widetilde{\omega} =
\omega - e \PhiH$, and the tortoise coordinate $\rho^*$, functions $R
(\rho)$ and $\Delta(\rho)$, and the potential $V^2$ are the same as those in
the section 2. Here we consider only the dominant mode
of $l = 0$. \\ \hspc
The wave equation (\ref{eqn:Radialdis}) is not invariant under rescaling
by the black hole mass $M$, in contrast to the case of the massless field
considered in the previous sections. The first and the last terms in the
bracket in Eq.(\ref{eqn:Radialdis}) are roughly proportional to $M^{-2}$,
whereas the second and the third terms are independent of the mass
scale. This results in that the transmission probability $|A|^2 - 1$
depends explicitly on the mass of the black hole. Hence we have to
calculate the emission rates for each mass scale and analyze the mass
dependence of the emission, in addition to the coupling constant
dependence. \\ \hspc
First we consider the Planck mass black hole $(M = M_{\mbox
{\scriptsize PL}})$. We show the emission rates in Fig.8 for four values
of the coupling constant: $\alpha = 0$, $0.5$, $1$, $1.5$. $Q$ is now
normalized by the mass of the black hole $M$, but not $\Qmax$, because
$Q$ itself is essential in this process, but not $Q/\Qmax$. We set the
particle mass $\mu = 0.001 M_{\mbox{\scriptsize PL}}$.
\begin{figure}
\begin{center}
\segmentfig{10cm}{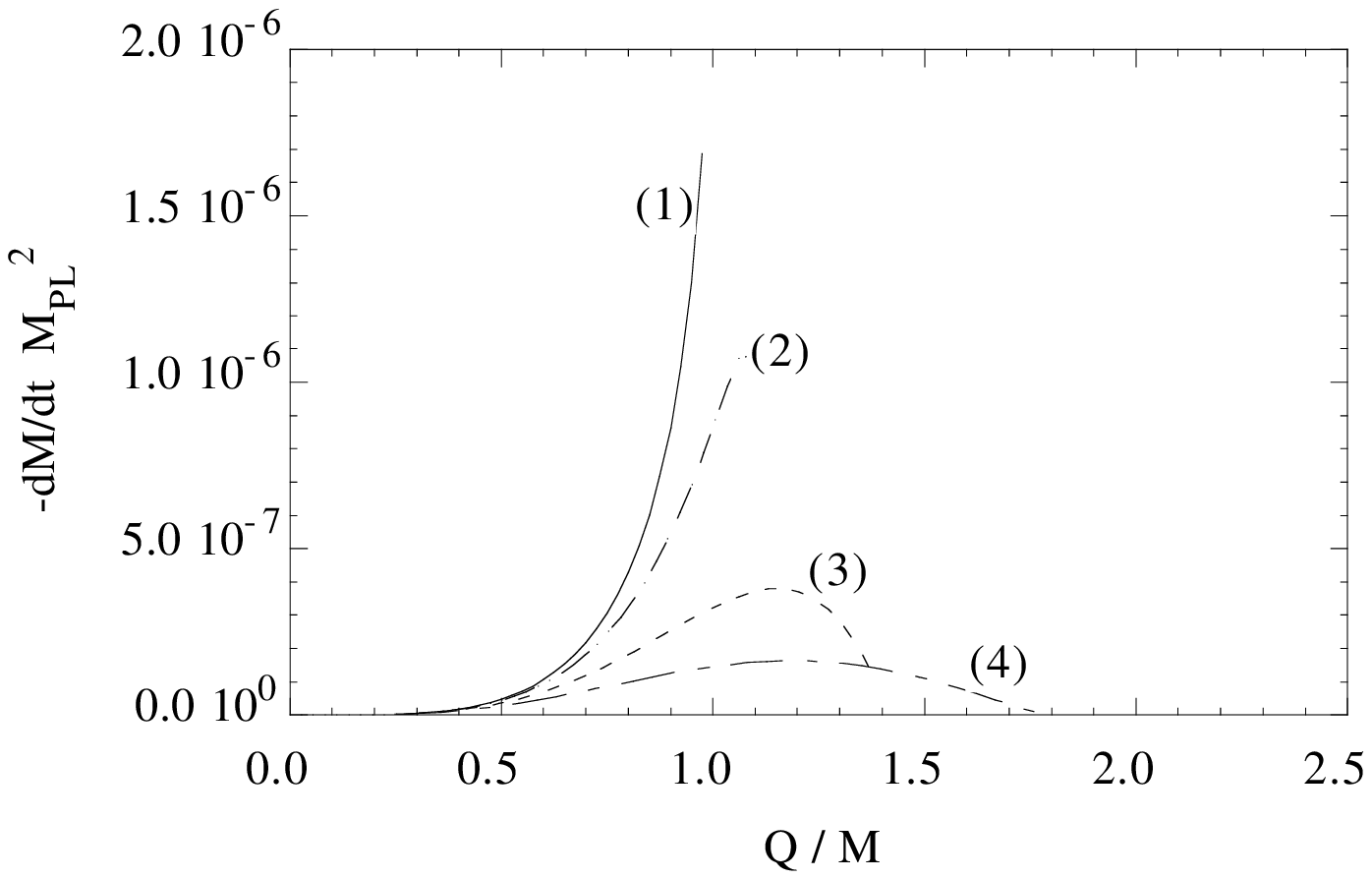}{(a)}
\vskip 1cm
\segmentfig{10cm}{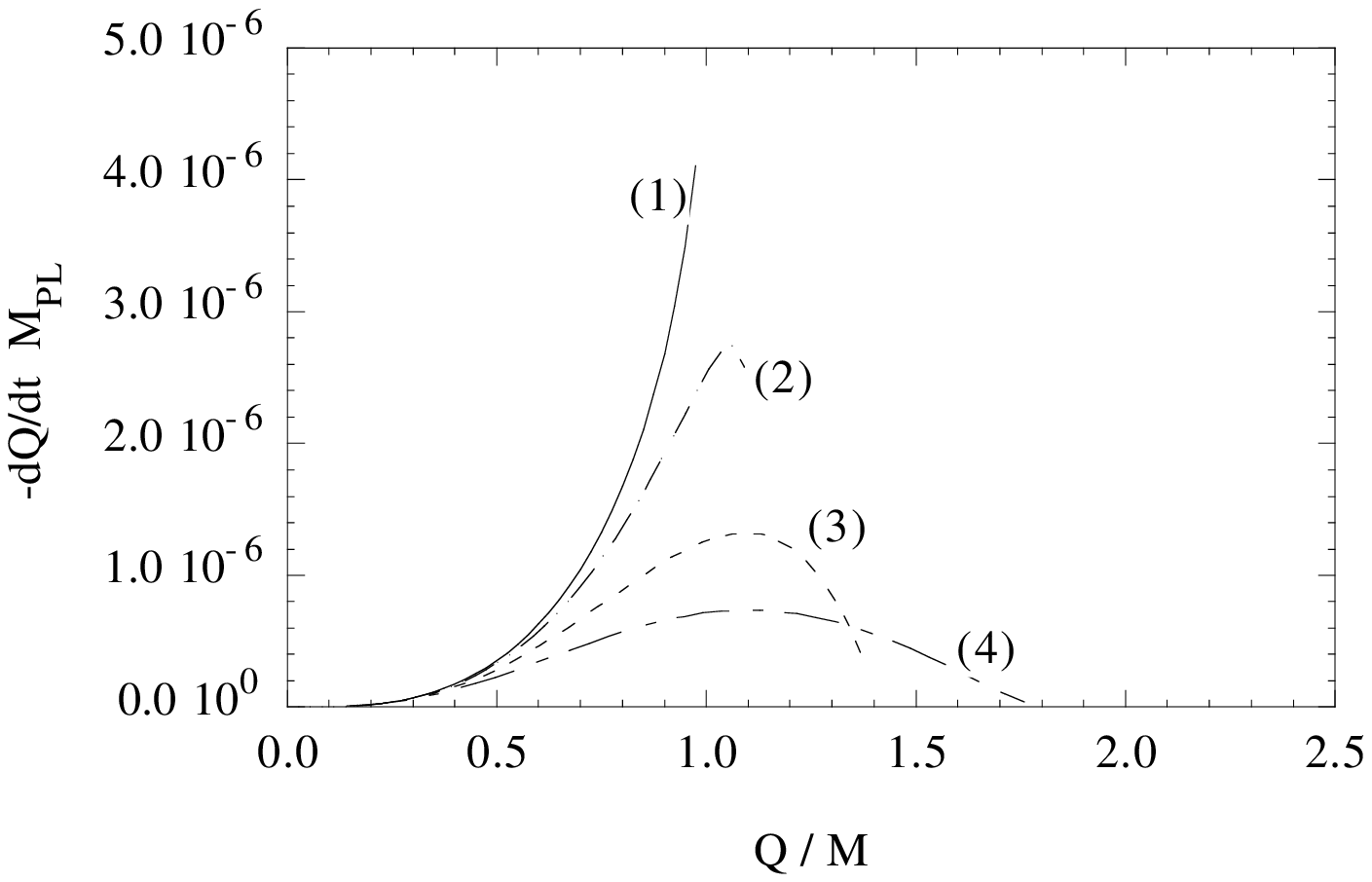}{(b)}
\end{center}
\begin{figcaption}{fig:discharge1}{15cm}
Discharge by superradiance from non-rotating black holes with
 mass $M = M_{\mbox{\scriptsize PL}.}$ \\
(a) and (b) show the energy emission rate $-dM/dt$, and the charge
emission rate $-dQ/dt$ normalized by the Planck mass $M_{\mbox
{\scriptsize PL}}$, respectively. Each line corresponds to (1):$\alpha =
0$, (2):$\alpha = 0.5$, (3):$\alpha = 1$, and (4):$\alpha = 1.5$.
\end{figcaption}
\end{figure}
{}From this figure, we find that the emission rates are greater
 in the black hole with
the smaller coupling constant, in contrast to the results of
the previous two sections. In particular, emission from the highly
charged black hole with larger coupling constant is very small. There
are two reasons for this. One is the behaviour of the electric potential
$\PhiH$. From Eq.(\ref{eqn:phiHdil}), we can see the electric potential
becomes smaller when the coupling constant $\alpha$ increases. The
second reason is that the effective potential in Eq.(\ref{eqn:Radialdis})
is very high near the extreme limit for the black hole with $\alpha > 1$
and the transmission probability becomes much smaller, as in the previous
cases.
\\ \hspc
Now we analyze the dependence of the emission rates on the mass of
the black hole. To see how the emission rate changes, we calculate the
case of $M = 10 M_{\mbox{\scriptsize PL}}$ and show the result in
Fig.9.
\begin{figure}
\begin{center}
\segmentfig{10cm}{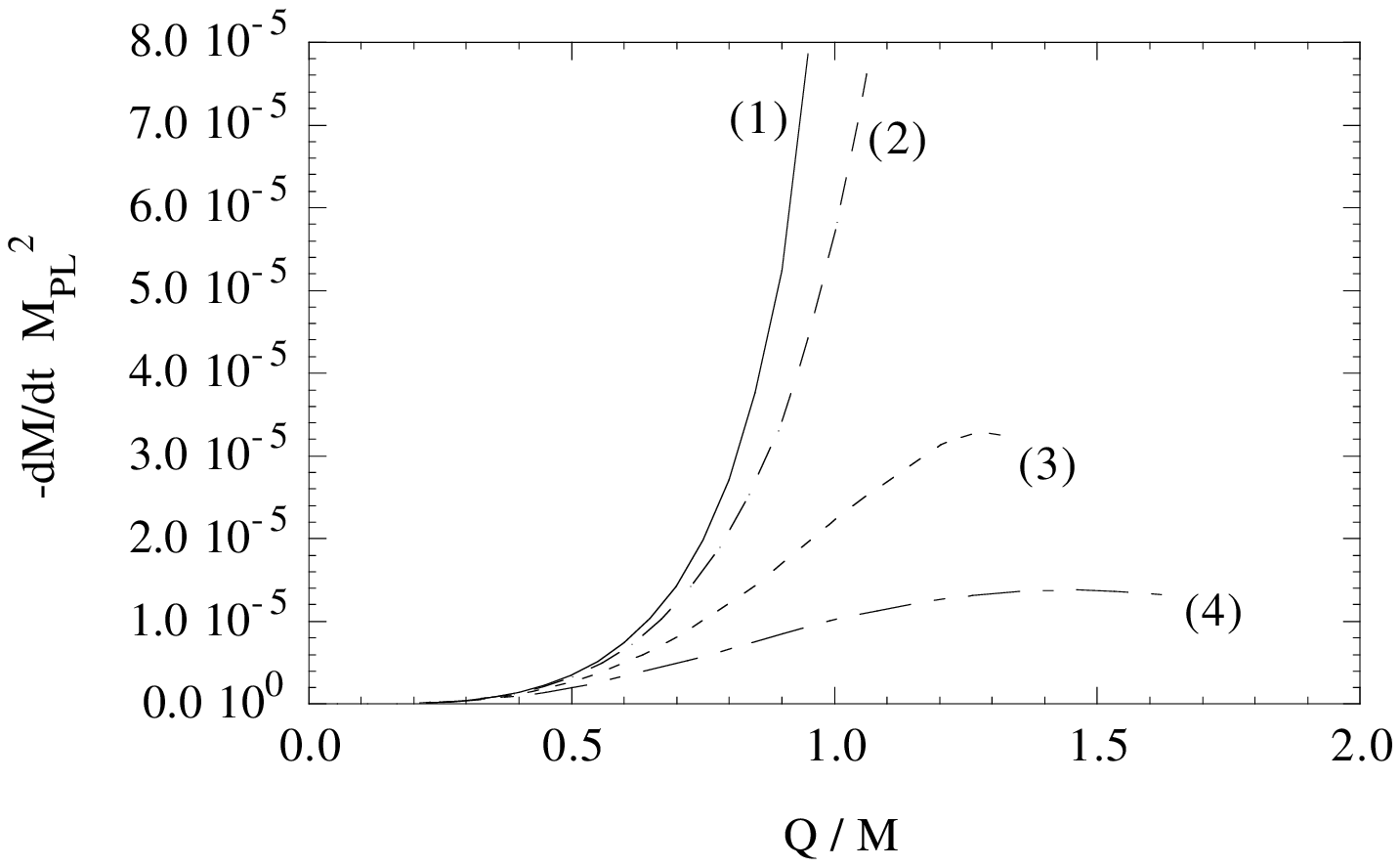}{(a)}
\vskip 1cm
\segmentfig{10cm}{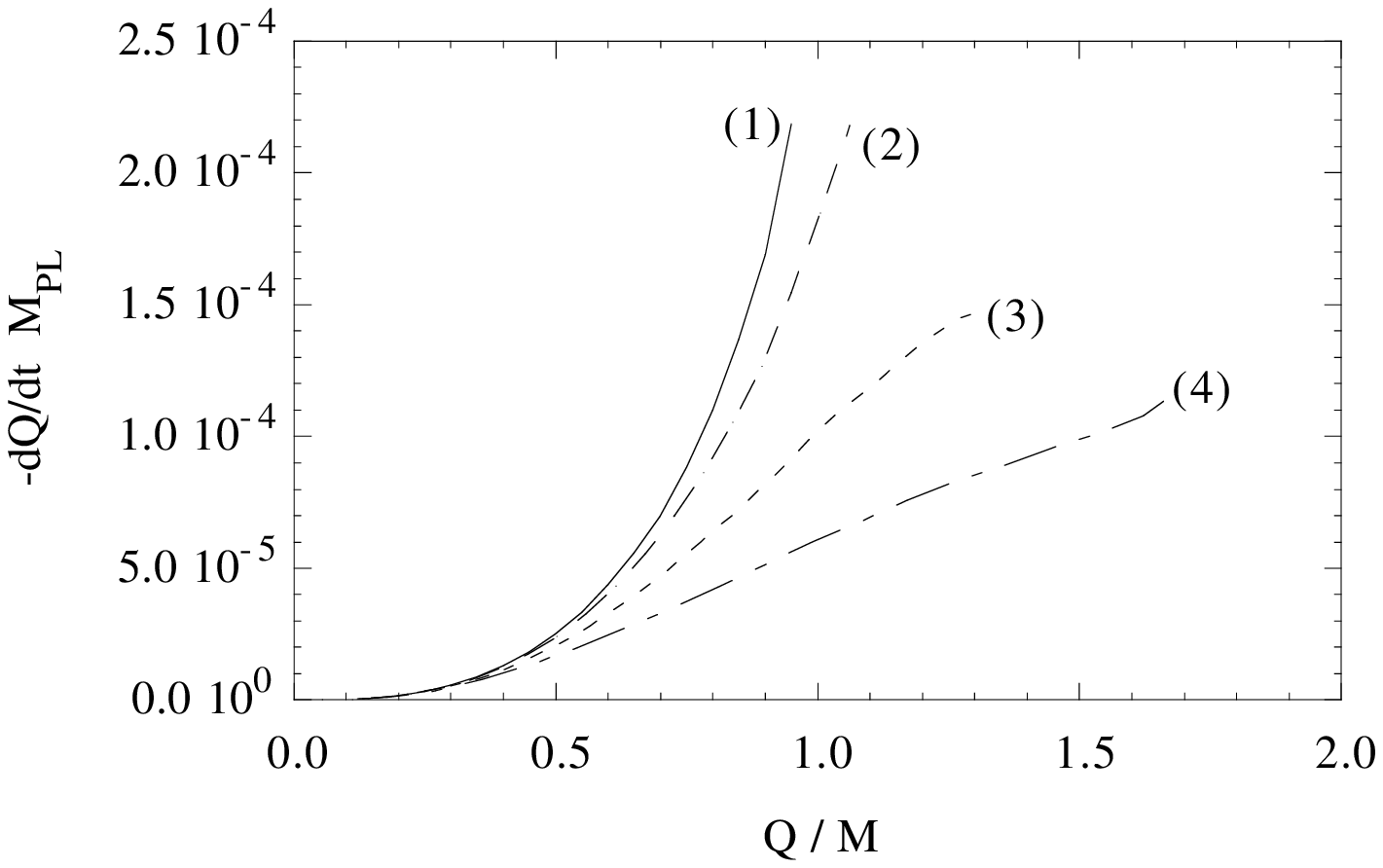}{(b)}
\end{center}
\begin{figcaption}{fig:discharge10}{15cm}
Discharge by superradiance from non-rotating black holes with
 mass $M = 10 M_{\mbox{\scriptsize PL}.}$ \\
(a) and (b) show the energy emission rate $-dM/dt$, and the charge
emission rate $-dQ/dt$, respectively. Each line corresponds to
(1):$\alpha = 0$, (2):$\alpha = 0.5$, (3):$\alpha = 1$, and (4):$\alpha =
1.5$.
\end{figcaption}
\end{figure}
Comparison with Fig.8 shows that the emission rates generally
 increase when the mass increases. This tendency is clearer in the highly
charged black holes with
larger coupling constant. The dependence on the coupling constant
is smaller than the case of $M = M_{\mbox{\scriptsize
PL}}$. This is because the height of the effective potential, which is
roughly proportional to $M^{-2}$, is effectively lower than that in the
case of a Planck mass black hole. In particular, the emission of a highly
charged black hole with large mass becomes insensitive to the coupling
 constant because the potential barrier gets small, compared with the
 case of a Planck mass-scale black
hole where the potential is very high for the large coupling constant
 and the emission is suppressed near the
extreme limit. Consequently, we expect that the coupling constant
dependence of the emission rate will become smaller as we increase the
mass of the black hole. \\ \hspc
For a black hole larger than $10 M_{\mbox{\scriptsize PL}}$, the
numerical calculation becomes difficult because we have to deal with a
very large scale black hole and a very small scale particle
simultaneously. Fortunately, for a massive black hole with small
charge, the $V^2$ term in Eq.(\ref{eqn:Radialdis}) is very small and can be
neglected. Furthermore, the rest of the potential terms (the second
and the third terms in the bracket) in Eq.(\ref{eqn:Radialdis}) vary very
slowly,
so we can use the W.K.B. approximation to calculate the transmission
probabilities, as in Ref.\cite{Shiraishi2,Gibbons}. \\ \hspc
When the black hole mass $M$ is sufficiently large and $Q/M$ is small,
the radial wave equation (\ref{eqn:Radialdis}) is approximated by
\begin{equation} \left[ \frac{\mbox{d}^2}{{\mbox{d} {\rho}^*}^{2}} +
\left( \omega - e \frac{Q}{\rho} \right)^{2} - \mu^2 \frac{\Delta(\rho)}
{R^2(\rho)} \right] \chi ({\rho}^*) = 0  \; ,
 \label{eqn:RadialWKB}
\end{equation}
and the transmission probability $|A|^2 - 1$ can be estimated from
\begin{eqnarray}
 |A|^2 - 1 & = & \exp\left[ - 2 \int^{\rf}_{\ri} \sqrt{|W|} \: \mbox{d}\rho^
* \right] \nonumber \\
& = & \exp\left[ - 2 \int^{\rho_{2}}_{\rho_{1}} \sqrt{|W|} \: \frac{R^2}
{\Delta} \: \mbox{d}\rho \right] \; ,
\label{eqn:WKBtrs}
\end{eqnarray}
where
\begin{equation}
 W = \left( \omega - e \frac{Q}{\rho} \right)^{2} - \mu^2 \frac{\Delta
(\rho)}{R^2(\rho)} \; ,
\end{equation}
and $\ri$ and $\rf$ ($\ri < \rf$) are the corresponding tortoise
coordinates to two roots $\rho_{1}$, $\rho_{2}$ of $W(\rho) = 0$. \\
\hspc
In the $\alpha = 0$ case, in which the black hole is described by the
Reissner--Nordstr\"om solution and
\begin{equation}
W = \left( \omega - e \frac{Q}{\rho} \right)^2 - \mu^2 \left( 1 - \frac{2
M}{\rho} + \frac{Q^2}{\rho^2} \right) \; ,
\end{equation}
Eq.(\ref{eqn:WKBtrs}) is integrated, giving
\begin{equation}
|A|^2 - 1 = \exp\left[ - 2 \pi \mu^2 \frac{e Q - \left( \omega - k \right)
M}{k \left( \omega + k \right)} \right] \; ,
\label{eqn:transRN}
\end{equation}
where
\begin{equation}
 k \equiv \sqrt{\omega^2 - \mu^2} \; .
\end{equation}
{}From this, we find Schwinger's formula for the emission rate
$\mbox{d}Q/\mbox{d}t$
\begin{equation}
\frac{\mbox{d} Q}{\mbox{d} t} \sim - \frac{e^4 Q^3}{\rho_{+}} \exp\left
[-\frac{\pi \mu^2 \rho_{+}^2}{e Q} \right]
\end{equation}
in the small charge limit\cite{Gibbons}. \\ \hspc
We can also explicitly calculate the transmission probability in
the superstring case ($\alpha = 1$), in which
\begin{equation}
W = \left( \omega - e \frac{Q}{\rho} \right)^2 - \mu^2 \left( 1 - \frac
{\rho_{+}}{\rho} \right) \; .
\end{equation}
It gives exactly the same result as Eq.(\ref{eqn:transRN}). In
addition, for the case of small charged black holes with
arbitrary coupling constant, we can use the approximation
\begin{equation}
\frac{Q}{\Qmax} \ll 1 \; ,
\end{equation}
so $\rho_{+} \gg \rho_{-}$, and then
\begin{equation}
\frac{\Delta(\rho)}{R^2(\rho)} = \left( 1 - \frac{\rho_{+}}{\rho} \right)
\left( 1 - \frac{\rho_{-}}{\rho} \right)^{(1 - \alpha^2)/(1 + \alpha^2)}
\sim \; \left( 1 - \frac{\rho_{+}}{\rho} \right) \left( 1 - \frac{\tilde
{\rho}_{-}}{\rho} \right) \; ,
\end{equation}
where
\begin{equation}
 \tilde{\rho}_{-} \: \equiv \:
 \frac{1 - \alpha^2}{1 + \alpha^2} \: \rho_{-} \; ,
\end{equation}
and we find the same transmission probability as Eq.(\ref
{eqn:transRN}). Hence, for the dilatonic black hole with a fixed mass
and charge, the transmission probability of the particle with the same
energy is hardly influenced by the coupling constant. As for the total
emission rate,  the black hole with the larger coupling constant emits a
little bit less energy, because the energy range of the superradiant
modes, i.e., $\mu \leq \omega \leq e \PhiH$, becomes narrow as the
coupling constant increases. When the charge of the black hole
increases, the emission rate increases. In the extreme limit, the black
 hole with larger
coupling constant can carry a larger charge. Hence we may
expect that the nearly extreme black hole with a larger coupling constant
emits larger energy than that with a smaller coupling constant. However,
near the extreme limit for $\alpha > 1$, the W.K.B. approximation breaks
down and the effective potential becomes very steep. As a
result, emission may not increase so much.
 So we expect that the
dependence of the emission on the coupling constant becomes smaller
for a more massive black hole. This has been confirmed by our numerical
calculations.

\section{Conclusion and Discussion}
\eqnum{0}
\hspc
In summary, we first studied the evaporation of dilatonic black
holes under the assumption that the black hole charge is conserved, and
analyzed its dependence on the dilaton coupling constant.
 We found that the
emission rate of the non-rotating black hole changes drastically at
$\alpha = 1$, which is the value predicted by superstring theory.
 In the case of
the coupling constant below unity, the emission rate vanishes in the
extreme limit, while the black hole with $\alpha > 1$ emits a large
amount of energy in the same limit, even though the potential barrier
becomes infinitely high in this case. This means the effect of the
temperature on the emission is stronger than that of the potential
barrier. \\ \hspc
As for rotating black holes, the temperature is zero for the
extreme black holes and the thermal emission also vanishes for all
known exact black hole solutions. However, in the maximally
charged limit $Q \rightarrow \Qmax$ of the Kaluza--Klein black hole,
while the angular momentum itself is still small, the angular
velocity of the black hole becomes very large and
 the effect of superradiance
becomes important. In superradiance, we also find the critical value of
the coupling constant $\alpha \sim 1$, above which the emission rate
increases rapidly as the black hole approaches the maximally
charged state. Therefore, we may reasonably conclude that $\alpha \sim 1$
is the critical coupling constant together with the thermal component
of the quantum radiation. \\ \hspc
As a result, a highly charged Kaluza--Klein black hole ($\alpha =
\sqrt{3}$) is inevitably accelerated towards evaporation into a naked
singularity. This situation is very similar to the final stage of the
evaporation of the Schwarzschild black hole where the emission blows
up and the area of the black hole vanishes. We expect that black
holes with $\alpha > 1$ show a similar evaporation process to the
Kaluza--Klein case, since the emission rates for such black holes are
very large in the maximally charged limit. \\ \hspc
We have also considered the discharge process by calculating
superradiance for non-rotating dilatonic black holes. If the mass of
the black hole is on the Planck scale, the emission is suppressed for large
coupling constants, compared with the Reissner--Nordstr\"om black hole
($\alpha = 0$), especially near the extreme limit. Hence, the effect of
the discharge may not be so important for highly charged black holes with
$\alpha > 1$. As the mass of the black hole increases, however, the
dependence of the emission on the coupling constant becomes small and a
black hole with any $\alpha$ will discharge efficiently. \\ \hspc
Holzhey and Wilczek\cite{HW} pointed out that, in the maximally
charged limit of the dilatonic black holes, the thermodynamical
interpretation breaks down. The solution of the maximally charged
limit represents a naked singularity, and the higher order quantum
effects will become important near this limit.
 This means that the black hole
thermodynamics may deviate from the conventional approach,
 which is based
on the semiclassical treatment of Hawking radiation. We should make
some comments on this point. The problems related to this paper are:
(1) The emission rate becomes very large, so we have to consider the
backreaction of the quantum effects on the metric, (2) The area of the
black hole vanishes in the maximally charged limit, which means we
have to deal with a horizon radius smaller than the Planck
scale, (3) To clarify the coupling constant dependence, we discuss
 the Planck mass-scale black hole. In order to study
such problems properly, we may need quantum gravity. However, before
investigating the full quantum theory, we first have to clarify
the
behaviour in the semiclassical regime.

\vspace{0.5cm}
{\bf Acknowledgment}\\
 We would like to thank R. Easther for reading the paper carefully.
This work was supported partially by the
Grant-in-Aid for Scientific Research  Fund of the
Ministry of Education, Science and Culture  (No.
06302021 and No. 06640412), and by Waseda University
Grant for Special Research Projects.

%

\vskip 2cm
\baselineskip .2in
\begin{thebibliography}{99}
\bibitem{Callan} C. G. Callan, D. Friedan, E. J. Martinec and M. J. Perry,
Nucl. Phys. B {\bf 262}  (1985)  593
\bibitem{GM} G. W. Gibbons and K. Maeda, Nucl. Phys.
B {\bf 298}  (1988)  741.
\bibitem{Dcos}  K. Maeda, Phys. Rev. D {\bf 35}  (1987)  471;  G.
Veneziano,  Phys. Lett. B {\bf 265}  (1991)  287;  D. S. Goldwirth and M.
J. Perry, Phys. Rev. D {\bf 49}  (1994)  5019; E. J. Copeland, A. Lahiri and
D. Wands, Phys. Rev. D {\bf 50}  (1994)  4868;  A. A. Tseytlin, Int. J. Mod.
Phys. D {\bf 1}  (1992)  223;  I. Antoniadis, J. Rizos and K. Tamvakis,
Nucl. Phys. B {\bf 415}  (1994)  497;  T. Damour and A. Vilenkin, Report
No. IHES/P/95/26  (unpublished)
\bibitem{unit}  We use the units of $c=\hbar=G=k_{\mbox{\tiny
B}}=1$.
\bibitem{GHS} D. Garfinkle, G. T. Horowitz and A.
Strominger, Phys. Rev. D {\bf 43}  (1991) 3140
\bibitem{HW} C. F. E. Holzhey and F. Wilczek, Nucl. Phys.
B {\bf 380}  (1992)  447
\bibitem{HH} J. H. Horne and G. T. Horowitz, Phys. Rev. D
{\bf 46}  (1992)  1340
\bibitem{Sen} A. Sen, Phys. Rev. Lett. {\bf 69}  (1992)  1006
\bibitem{KM} J. Koga  and  K. Maeda,  Phys. Lett. B {\bf 340}
(1994)  29
\bibitem{FZB} V. Frolov, A. Zelnikov and U. Bleyer, Ann.
Phys. (Leipzig) {\bf 44}  (1987)  371
\bibitem{Shiraishi1} K. Shiraishi, Phys. Lett. A {\bf 166}
(1992) 298
\bibitem{GW} G. W. Gibbons and D. L. Wiltshire, Ann.
Phys. (N. Y.) {\bf 167}  (1986)  201
\bibitem{Shiraishi2} The coupling of dilaton with a scalar field
 may change our results.
 For example, Shiraishi analyzed the superradiant
discharge process of such a scalar field in the
spherically symmetric dilatonic black hole and showed
 that the emission
rate is enhanced when $\alpha$ increases, due to the coupling between
dilaton and scalar field, in contrast to our results.
 See  K. Shiraishi, Mod. Phys. Lett. A {\bf 7}
(1992)  3449\\
\bibitem{Hawking} S. W. Hawking, Commun. Math. Phys. {\bf 43}
(1975)  199
\bibitem{Bouwkamp} C. J. Bouwkamp,  Phillips. Res. Rep. {\bf 5}
(1950)  87
\bibitem{Unruh} W. G. Unruh, Phys. Rev. D {\bf 10}  (1974) 3194
\bibitem{Gibbons} G. W. Gibbons,  Commun. Math. Phys.
{\bf 44}  (1975)  245
\end{thebibliography}
\end{document}